\newcommand{\RHd}{Quantifying QN excitations}

\documentstyle[aps,epsf]{revtex}

\begin{document}

\preprint{Quantifying QN excitations}

\title{
Quantifying excitations of quasinormal mode systems}
\author{Hans-Peter Nollert}
\address{
Astronomy and Astrophysics, University of T\"{u}bingen, 72076 T\"{u}bingen, Germany}
\author{Richard H. Price}
\address{
Department of Physics, University of Utah, Salt Lake City, UT 84112}

\maketitle

\newcommand{\mpb}[1]{\vbox{\hsize=2.truecm\raggedright #1}}
\newcommand{\TDP}{{TDP}}
\newcommand{\sTDP}{{spiked TDP}}
\newcommand{\Lpsi}{\hat\psi}
\newcommand{\running}

\pagestyle{myheadings}

\bigskip

%

  Computations of the strong field generation of gravitational waves
  by black hole processes produce waveforms that are dominated by
  quasinormal (QN) ringing, a damped oscillation characteristic of the
  black hole. We describe here the mathematical problem of quantifying
  the QN content of the waveforms generated.  This is done in several
  steps: (i) We develop the mathematics of QN systems that are
  complete (in a sense to be defined) and show that there is a
  quantity, the ``excitation coefficient,'' that appears to have the
  properties needed to quantify QN content. (ii) We show that
  incomplete systems can (at least sometimes) be converted to
  physically equivalent complete systems.  Most notably, we give a
  rigorous proof of completeness for a specific modified model
  problem.  (iii) We evaluate the excitation coefficient for the model
  problem, and demonstrate that the excitation coefficient is of
  limited utility. We finish by discussing the general question of
  quantification of QN excitations, and offer a few speculations about
  unavoidable differences between normal mode and QN systems.


\bigskip
\pacs{PACS numbers: 02.10.Sp, 02.30.Lt, 04.30.Db, 04.70. s}
PACS numbers: 02.10.Sp, 02.30.Lt, 04.30.Db, 04.70. s

\pagebreak

%
\renewcommand{\RHd}{Quantifying QN excitations}  \markboth{\RHd}{\RHd}

\section{Introduction and Outline}

Essentially all computations of the generation of gravitational waves
by strong field black hole processes produce a gravitational wave with
the shape of a damped sinusoid\cite{BHQNrefs}. The oscillation period
and damping time depend only on the parameters of the black hole, and
not on the manner of excitation. The meaning of the complex frequency
of this damped oscillation is now well understood. A single frequency
perturbation outside the hole can satisfy the natural radiative
boundary conditions (radiation into the black hole and outward to
infinity) only if the frequency is one of the discrete set of
frequencies, called quasinormal (hereafter QN) frequencies. The least
damped of these complex frequencies is what dominates the appearance
of computed waveforms.

QN excitations are relevant, in principle, to most or all systems with
radiative boundary conditions. Stellar models for example have short
periods for nonradial oscillations driven by fluid pressures, and long
damping times of these fluid oscillations due to the weak emission of
gravitational waves.  The motions of the stellar fluid can be studied
with radiation damping omitted (e.g., with the use of Newtonian
gravitation theory, or Post-Newtonian theory) and the weak
radiation can be added, after the fact.  When the radiative coupling
is ``turned off'' the problem of the oscillation of a perfect fluid
stellar model can be analyzed in normal modes\cite{beyer,beyerschmidt}
and one can find the radiated energy coming from each separate
oscillation frequency, and can decompose the radiative power into that
fraction assigned to each frequency.

The situation is dramatically different for black holes, which have
only a single time scale. (For a nonrotating hole this is $2GM/c^3$
where $G$ is the universal gravitational constant, $c$ is the speed of
light, and $M$ is the mass of the hole.) The period and damping time
are therefore of the same order and there is no meaningful way of
turning off the damping for black hole oscillations; there is no
underlying normal mode system.  This suggests that there may be no
clear way of specifying ``how much QN ringing'' of some particular
black hole QN frequency is contained in an emitted waveform. This
suggestion is made plausible by the mathematical origins of normal
modes and QN modes. The properties and usefulness of normal modes are
closely related to the fact that they are eigensolutions to a self
adjoint problem. QN modes, on the other hand, are
eigensolutions of a problem that is not self adjoint.
But the dominance of QN frequencies in computed waveforms is so robust
that it seems that the strength of QN ringing {\em must} be
quantifiable, or at least that mathematical sense must be made of the
question.

In this paper we try to make mathematical sense of quantification. In
attempting this we draw upon parallels with normal modes systems. By
the ``excitation'' of a mode we mean, in parallel to excitation in
normal modes systems, an index of the contribution that each
mode makes to the overall waveform and to the energy.
To develop a description of QN excitation we start with a viewpoint
that a meaningful and rigorous quantification is very implausible
unless the QN system is, in some sense, complete.  We then follow a
three step process.  First, in Sec.~II, we define and posit the
existence of QN systems that are complete (in a sense to be defined).
We then point out difficulties in quantifying excitation in a complete
QN system.  We construct a particular measure, the ``excitation
coefficient'' that overcomes these difficulties, and is closely
related to the description of the excitation of normal modes.

Our next step is to prove the existence of complete QN systems and
relate the mathematics of black hole processes to complete systems.
This step, carried out in Sec.\,III requires a rather lengthy
discussion of ``induced completeness.'' Though this discussion is not
directly related to the problem of QN excitation, it is a necessary
step (and is interesting in its own right).  The discussion in
Sec.\,III shows that completeness can be induced.  That is, an
incomplete QN system can be changed with a modification that satisfies
two criteria: (i) The effect of the modification can be made
arbitrarily weak. More specifically, the modification can be made
small enough so that the waveform that evolves from any initial
conditions is arbitrarily close to the waveform evolving with no
modification.  (ii) No matter how weak the modification is, the
modified QN system is complete.  Our demonstration in Sec.\,III does
not consists of a general theory for such modifications; a conjecture
about the general conditions has been given by Young et
al.\cite{youngsuenetal} (though their definition of completeness is
somewhat different from ours). Here we will sacrifice generality 
and direct astrophysical relevance for specificity and rigor.
We present the details of a specific model. We will start with
a model, the '\TDP' with only a single conjugate pair of QN
frequencies, and modify it to the ``\sTDP,'' a model with an infinite
QN spectrum.  The Appendix gives a rigorous proof of completeness of
the \sTDP (i.e., that under specified circumstances the outgoing
waveform is a convergent sum of components at quasinormal
frequencies).  Numerical results are shown in Sec.\,III to demonstrate
the negligible effect of the modification, and to demonstrate the
pattern of convergence of sums of single frequency excitations.

Having established that completeness can be induced (at least in one
model problem), we return, in Sec.\,IV, to the question of measuring
the excitation of QN modes
and in particular to the excitation
coefficient, introduced in Sec.\,II.  We demonstrate, with a few
examples, that this formal measure of excitation does not generally
give a useful quantification. The failure of this measure is
discussed along with a broader discussion of differences between QN
and normal mode systems, and conjectures about mathematical properties
of QN systems.

%

\renewcommand{\RHd}{Quantifying QN excitations}  \markboth{\RHd}{\RHd}

\section{Complete QN systems} 
\subsection{Definition of QN frequencies}
For definiteness we will limit considerations to solutions of the 
equation
\begin{equation}
  \label{basiceq}
\frac{\partial^2 \Psi}{\partial x^2}-\frac{\partial^2 \Psi}{\partial t^2}
-V(x)\Psi=0\ .
\end{equation}
Such an equation describes the dynamics of many mechanical systems,
and the evolution of multipole perturbations (scalar, electromagnetic,
or gravitational) of spherically symmetric (Schwarzschild) black holes\cite{RW,zerilli,Chandrabook}. Perturbations of rotating (Kerr)
holes\cite{Chandrabook}, on the other hand, cannot be reduced to
radial-time equations.  QN oscillations are single frequency solutions
of the form $\Psi(t,r)=\psi(x)\exp{(i\omega t)}$ and hence are
solutions of the equation
\begin{equation}
  \label{eigeneq}
  \frac{\partial^2 \Psi}{\partial x^2}+\left[\omega^2
-V(x)\right]\Psi=0\ .
\end{equation}

We assume that the domain of $x$ includes $x=\infty$, and that the
nature of the potential $V(x)$ is such that a ``radiative boundary
condition'' can be defined at $x\rightarrow\infty$.  A clear example
is a potential with support of $V(x)$ only for $x$ less than some
$x_{max}$.  In this case the boundary condition is that
$\psi(x)\propto \exp{(-i\omega x)}$ for $x>x_{max}$. 

For potentials that do not vanish, but fall off sufficiently fast as
$x\rightarrow\infty$, the more general radiative boundary condition
for real frequencies will be that for large $x$, the solution for
$\psi(x)$ have the form $\exp{(-i\omega x)}F(\omega,x)$, with
$F\rightarrow$\,constant, as $x\rightarrow\infty$. This condition is
not quite sufficient if $\omega$ has a positive imaginary part; see
\cite{NollertSchmidt} for a complete discussion. In short, the
solution satisfying a radiative boundary condition for complex
$\omega$ can be regarded as an analytical continuation of a solution
satisfying a radiative boundary condition for real $\omega$.

The boundary condition at the other end  of of the $x$ domain may be a
standard Sturm-Liouville boundary condition (e.g., $\Psi=0$ at $x=0$)
or may be a radiative boundary condition at $x\rightarrow-\infty$. For
spherically symmetric black holes, the range of $x$ extends from
$-\infty$ to $\infty$, with $-\infty$ representing the black hole
horizon\cite{NollertSchmidt,Leaver}.  The potentials fall off exponentially in $x$ as
$x\rightarrow-\infty$ and as ${\rm const}/x^2$, as
$x\rightarrow\infty$. Radiative boundary conditions are imposed both
at $x\rightarrow -\infty$ and $x\rightarrow\infty$, corresponding to
radiation moving inward through the black hole horizon, and outward
towards spatial infinity.

QN frequencies, are the eigenvalues $\omega=\omega_{QN}$ to the
problem defined by (\ref{eigeneq}) for radiative boundary conditions
of the type just discussed. Due to the boundary conditions, this
problem is generally not of the Sturm-Liouville type and the usual
features of eigenvalues of a Sturm-Liouville problem are absent.  In
particular the QN frequencies $\omega_{QN}$ are generally not real.  A
positive imaginary part indicates an exponential decrease with time.
A negative imaginary part would indicate an instability; no
frequencies with negative imaginary parts have been found for black
hole QN systems.

It is clear that QN frequencies must occur in conjugate pairs. If
$\omega_{QN}$ is a solution to the eigenproblem corresponding to
$\psi_{QN}$, then $-\omega_{QN}^*$ is also a solution corresponding to
$\psi_{QN}^*$.  We will use a tilde ($\tilde{\ }$) to denote the
conjugate relationship of QN frequencies. Thus the conjugate to QN
frequency $\omega_7$ is $\omega_{\tilde{7}}$, that is
$\omega_{\tilde{7}}=-\omega_7^*$.

\subsection{Definition of completeness}

We will choose to give a rather specific meaning to a complete system
of QN modes.  The Cauchy data for (\ref{basiceq}) consists of
$\psi_0$ and $\dot{\psi}_0$, the initial value of $\Psi$ and of its
time derivative,
\begin{equation}
  \label{cauchysymbs}
  \psi_0(x)\equiv\left.\Psi(t,x)\right|_{t=0}\ \ \ 
  \dot{\psi}_0(x)\equiv\left.\partial\Psi(t,x)/\partial t\right|_{t=0}\ \ \ 
\end{equation}
We consider an interval $x_2<x<x_1$ in
the domain of $x$ and we consider  Cauchy data at $t=0$ for
(\ref{basiceq}) which has support only in this interval. We then
consider the solutions $\Psi$ to (\ref{basiceq}) for such data, and 
we focus attention on the value of this solution at $x_{\rm
obs}$, a particular value of $x$ satisfying $x_{\rm obs}>x_1$. This
corresponds to the physical situation of an observer at $x_{\rm
obs}$ detecting radiation resulting from an initial disturbance
located at some distance from her. The ``observed waveform'' that we
focus on is then
\begin{equation}
f(t)\equiv\Psi(x_{\rm obs},t)\ .
\end{equation}
\begin{figure}
\hspace*{120pt}\epsfxsize=150pt \epsfbox{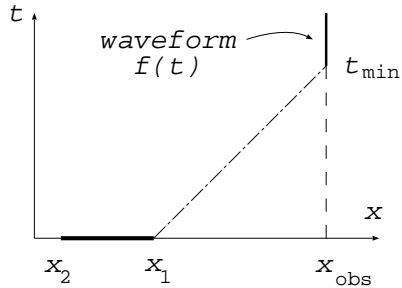}
\caption{\label{fig:observ}
Propagation of initial data to the observation location
$x_{\rm obs}$ defining the waveform $f(t)$.}
\end{figure}
As shown in Fig.~\ref{fig:observ}, 
there will in general be a minimum value $t_{\rm
min}$ of $t$, such that the point $x_{\rm obs},t$ is influenced by the
evolved Cauchy data. That is, for $t<t_{\rm min}$, the area between the past
directed characteristics from $t,x_{\rm obs}$ intersects the $t=0$
hypersurface outside the support of the Cauchy data.
We are interested in $f(t)$ only for $t\geq t_{\rm min}$.  The
physical interpretation of this is that we are considering only the
waveform generated by the Cauchy data. 

We take a complete QN system to be one which satisfies the following
criteria:
\begin{itemize}
\item The solutions to the QN eigenvalue problem form a discrete
spectrum and can be arranged in order of 
increasing $|\Re(\omega_n)|$. 
\item We consider only  Cauchy data that
\begin{description}
\item[--] has support only within a compact region $[x_2,x_1]$.
\item[--] belongs to a specific continuity class $C^p$, where $p$ 
          depends on the nature of the problem.
\item[--] results in a wave form which is square integrable from 
          $t=t_{\rm min}$ to $\infty$.
          
\end{description}
\item For such Cauchy data, the waveform $f(t)$ that evolves
          from any such allowed Cauchy data can be written as
\begin{equation}\label{compdef}
f(t)=\sum_{n} a^n\,e^{i\omega_n t}\ .
\end{equation}
Here $a^n$ is the n$^{\rm th}$ coefficient in the sum over QN modes.
Since $f(t)$ is a function of 
$x_{\rm obs}$, the 
$a^n$ coefficients are also functions of  $x_{\rm
  obs}$, but we shall not explicitly exhibit this dependence.  The
summation in (\ref{compdef}) is in order of increasing 
$|\Re(\omega_n)|$, and the convergence is 
uniform for $t>t_{\rm
  min}$.
\end{itemize}
It is important to note that our view of completeness is rather
different from other possible meanings of the term. In particular, our
choice of the meaning of completeness has nothing directly to do with
the $x$-dependence of the single frequency solutions and with the
question of whether these solutions can be used to span acceptable Cauchy
data. Our meaning of completeness, then, is rather different from that
of Young et al.\cite{youngsuenetal}. It also disagrees with the
concept of 
completeness used by
Husain and Price\cite{husainprice} and by Beyer\cite{beyer}, and Beyer
and Schmidt\cite{beyerschmidt}

\subsection{Function space and inner product}\label{InnerProd}
In accordance with our definition of completeness, our vector space is
the space of all functions $f(t), t\geq t_{\rm min}$ that can evolve
from acceptable Cauchy data. Our class of acceptable Cauchy data will
always be chosen so that $f(t)$ is square integrable from $t=t_{\rm
  min}$ to $\infty$.  On this space of functions we define an inner
product to be:
\begin{equation}\label{dotdef}
f\cdot g\equiv\int_{t_{\rm min}}^\infty f^*(t)g(t)\,dt.
\end{equation}
We could of course include a weight function  $W(t-t_{\rm min})$
in the integral defining the inner product, but the time translational
symmetry of the background suggests that $W$ should be constant. 
The choice in Eq.\,(\ref{dotdef}), furthermore, means that $f\cdot f$ 
is the time integral of the square of the wave function, a measure closely
related to the energy content of a wave. (For black hole processes,
the connection with gravitational wave power will be made explicit presently.)
Our assumption of completeness above means that the functions
$\exp{(i\omega_{QN} t)}$, while
not elements of our function space themselves, span this function
space in the sense of (\ref{compdef}); we will therefore consider them
a basis. For a 
function $f(t)$ in
our space we can use the inner product to compute another set of
coefficients $a_n$ by
\begin{equation}\label{covdef}
a_n\equiv (e^{i\omega_n t})\cdot f(t)=
\int_{t_{\rm min}}^\infty e^{-i\omega^*_n t}f(t)\,dt\ .
\end{equation}
The following relations for the coefficients of conjugate modes are straightforward to verify:
\begin{equation}
  \label{conjcoeffs}
  a_{\tilde{k}}=(a_k)^*\ \ \ \ \   a^{\tilde{k}}=(a^k)^*\ \ \ \ \ 
\end{equation}

Since the convergence is uniform by hypothesis, we can integrate term
by term in the sum-of-modes expression for the norm of $f$ to find:
\begin{equation}
  \label{sumofsq}
\int_{t_{\rm min}}^\infty|f(t)|^2dt=\int_{t_{\rm min}}^\infty \left(\sum_{n}a^n\,e^{i\omega_n t}\right)^*f(t)\,dt=
  \sum_{n}(a^n)^*a_n\ .
\end{equation}
The final sum in (\ref{sumofsq}) is real, as it must be, since for any
$k$, the sum $(a^k)^*a_k+$ $(a^{\tilde{k}})^*a_{\tilde{k}}$ is real.

In most physical problems the radiated power is the square of the time
derivative of the waveform. If at $x_{\rm obs}$ our wave function
evolving from the initial data is $f(t)\equiv\psi(x_{\rm obs},t)$.
If $f(t_{\rm min})$ vanishes (i.e., if the waveform starts continuously) 
then this type of energy can be evaluated as
\begin{equation}\label{energysum}
\int_{t_{\rm min}}^\infty|\dot{f(t)}|^2dt=\sum_{n}\left(\omega_n^*\right)^2(a^n)^*a_n\ .
\end{equation}
As in (\ref{sumofsq}) the reality of the sum is guaranteed by the 
relations of conjugate coefficients in (\ref{conjcoeffs}).

Since we have an inner product, we have an equivalence between vectors
and dual vectors in our function space and we can
define a set of covariant basis functions $\phi^m(t)$ by the
property $\phi^m(t)\cdot e^{i\omega_nt}=\delta_{nm}$.  If follows from
Eq.\,(\ref{covdef}) that the $a_n$ are the expansion coefficients for
$f(t)$ with respect to the covarariant basis functions $\phi^n$. We
shall henceforth refer to $a^n$ and $a_n$, respectively, as the
contravariant and covariant coefficients of $f(t)$.  The components of
the metric, in this function space, with respect to the QN basis, are
\begin{equation}\label{metricdef}
\left(e^{i\omega_n t}\right)\cdot\left(e^{i\omega_k t}\right)
=\int_{t_{\rm min}}^\infty e^{it(\omega_k-\omega^*_n)}dt\equiv G_{nk}\ .
\end{equation}
It should be noted that $G$ is a hermitian matrix, but it is not
diagonal, i.e., the QN oscillations are not orthogonal.  The metric
coefficients can be used, in principle, to relate $a^n$ and $a_n$.
The expression for $f(t)$ in Eq.\,(\ref{compdef}) can be substituted
in Eq.\,(\ref{covdef}). Since the convergence in Eq.\,(\ref{compdef})
is uniform, we can integrate term by term and get
\begin{equation}\label{covfromG}
a_n=\sum_kG_{nk}a^k\ .
\end{equation}
Since $G_{nk}$ is not diagonal, the covariant basis vectors, i.e., the
basis vectors dual to $e^{i\omega_nt}$, are mixtures of the
$e^{i\omega_nt}$ functions (usually involving all of them).
An indication of the unfamiliar problems this produces can be seen in
the following rough argument.  Let us suppose that we have a waveform
that in some sense is ``almost pure'' (say) seventh mode.  That is,
suppose that $f(t)\approx
a^7e^{i\omega_7t}+a^{\tilde{7}}e^{i\omega_{\tilde{7}}t}$.  (We are
supposing that this relationship is only approximate since it will in
general be impossible to excite a truly pure single frequency mode
with smooth, compact initial data.) This waveform, for which (almost)
the only contravariant coefficients are $a^7$ and $a^{\tilde{7}}$, will
have contravariant coefficients $a^n$ for all $n$. A waveform that is a pure
(or almost pure) single mode excitation in one sense is therefore not
a single mode excitation in another. This presages some of the
problems in quantifying the excitation of a mode, and we will return
to this point at the end of Sec.\,IV.

It is possible, of course, to use Gram-Schmidt orthogonalization to
find basis function which are orthogonal according to the inner
product of (\ref{dotdef}). The resulting basis functions will not
(except for one of them) correspond to single frequency excitations, and
do not seem to be of interest.

\subsection{Intuitive insights; the excitation coefficient}\label{Intuit}

\begin{figure}
\hspace*{30pt}\epsfxsize=230pt \epsfbox{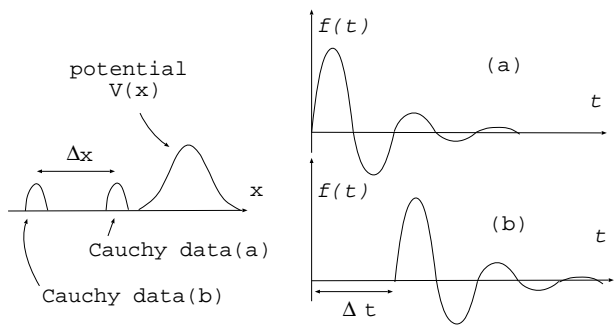}
\caption{\label{fig:shift}
Initial data shifted in location produces identical waveforms
shifted in time.}
\end{figure}

Some rough considerations of model problems suggest the intuitive
basis for some of the mathematical difficulties to appear below, and
point to a possible approach to quantifying excitation.  The most
obvious difficulty is the `time shift' problem. We imagine a
configuration like that pictured in Fig. (\ref{fig:shift}): A
potential with compact support and two sets of Cauchy data $(a)$ and
$(b)$. The two Cauchy data sets are localized and are identical except
that set $(b)$ is shifted to the left, to smaller $x$, by some finite
displacement $\Delta x$. The support of neither Cauchy data set
overlaps the support of the potential. In this case it is clear that
the waveform generated from the two cases will be identical except
that the waveform from $(b)$ will be shifted to later times, relative
to that for $(a)$, by an amount $\Delta t=\Delta x$.  Each
contravariant coefficient will therefore be larger in the $(b)$
waveform than in the $(a)$ waveform.  The the conjugate pair of
contravariant coefficients $\{a^7,a^{\tilde{7}}\}$ will, for example,
be larger by $\exp{[(\Im{\omega_7})\Delta t]}$ for $(b)$ than for
$(a)$, though the excitation is physically identical.  The analogous
difficulty does not arise for normal modes; since they are not damped,
the time delay only causes a phase shift.  The trend is opposite for
the covariant coefficients $\{a_7,a_{\tilde{7}}\}$. As defined in
(\ref{covdef}) $\{a_7,a_{\tilde{7}}\}$ will be smaller for $(b)$,
since $\exp{(-i\omega_7^* t)}$ (or $\exp{(-i\omega_{\tilde{7}}^* t)}$)
are smaller by $\exp{[(-\Im{\omega_7})\Delta t]}$ at the later times
during which the $(b)$ waveform has support.  Therefore, neither the
coefficients $a^k$ nor $a_k$ alone can provide a useful measure for
the excitation of a QN mode.

Any useful measure of excitation {\em must} give the same result for
the two waveforms in Fig. (\ref{fig:shift}). We take advantage of the
opposite tendencies of the contravariant and covariant coefficients
under time shift to define a quantity that is the same for both
waveforms, the ``excitation coefficient'' $A_k$ for the $k^{\rm th}$
QN mode:
\begin{equation}
  \label{excoeffdef}
  A_k\equiv(a_k)^*a^k 
+(a_{\tilde{k}})^*a^{\tilde{k}} \ .
\end{equation}
We conjecture that these excitation coefficients, or quantities constructed
from them (sums of excitation coefficients, functions of excitation
coefficients, etc.) or quantities very closely related (see the energy
excitation coefficient, below) are the only relevant mathematical
objects in the vector space that are unaffected by the time shifts,
and have the additional properties that we outline in the following.

In addition to the insensitivity of the excitation coefficient to 
time shifts of waveforms, the excitation coefficient has another important 
and relevant property (cf. Eq.~(\ref{sumofsq})):
\begin{equation}\label{sumofexcoeffs}
\int_{t_{\rm min}}^\infty|f(t)|^2dt= \sum_{k}A_k
\ ,
\end{equation}
where the sum is over conjugate pairs. The excitation coefficients
of the complete set of QN modes sums to the norm of the waveform. We can
define a quantity  closely related to $A_k$:
\begin{equation}
  \label{defenexcoeff}
 E_k\equiv(\omega_k^*)^2(a_k)^*a^k 
+(\omega_k)^2(a_{\tilde{k}})^*a^{\tilde{k}} \ .
  \end{equation}
  We shall refer to the $E_k$ as the energy excitation coefficient.
  According to (\ref{energysum}), the sum over conjugate pairs of
  these coefficients gives the norm of $\dot{f}$.  The summation
  properties of the $A_k$ and the $E_k$ are important in clarifying
  what we mean by the ``excitation'' we are attempting to quantify.
  These coefficients appear to tell us the contribution made
  by each mode to a measure of the waveform. In the case of black hole
  perturbations it turns out that it is possible to make an even more
  direct connection. If $f(t)$ is a solution of the Zerilli
  equation\cite{zerilli} for even parity perturbations, then the
  integral on the left of (\ref{energysum}) is proportional to the
  radiated energy and $E_k$ has the appearance of the energy in the
  k$^{\rm th}$ mode.  If $f(t)$ is a solution of the Regge-Wheeler
  equation\cite{RW} for odd parity perturbations, then the integral in
  Eq.~(\ref{sumofsq}) is the energy and $A_k$ has the appearance of
  the energy in the k$^{\rm th}$ more.

The possibility of quantification with the excitation coefficient
(or energy  excitation coefficient) will be a central focus, of 
Sec.\,IV, but before we begin specific computations, there are
a few more possibly useful insights that can be found from intuitive 
considerations. For one thing, it is interesting that the two sets of
coefficients can be related to 
two different aspects of an emitted wave. The contravariant
coefficients, telling us how much of a certain mode must be added in
order to get the waveform, can be considered a ``theoretical''
coefficient. For a given waveform, the projection operation on the
waveform defined by (\ref{covdef}) can be considered to give the
``experimental'' excitation coefficient.

Simple considerations can also produce insights into the nature of the
metric matrix $G_{nk}$ defined in (\ref{metricdef}). 
Consider the signal produced by some Cauchy data, and let the vector
$\bar a$ denote the contravariant coefficients of the resulting wave
form with respect to the QN mode basis. Let the vector of covariant
coefficients be given by $\underline a$, so that 
$\underline a = G \bar a$. Now consider the same Cauchy data, shifted
to the left by $\Delta x$. The new contravariant and covariant
coefficient vectors will be given by: 
$\tilde{\bar a}^k = \exp{(-i\omega_k\Delta x)} \; {\bar a}^k$ and
$\tilde{\underline a}_k = \exp{i\omega_k)\Delta x)} \; {\underline a}_k$.
Taking the standard linear algebra norm,, we have 
$||\;\tilde{\bar a}\;|| = \alpha \; ||\;{\bar a}\;||$ and 
$||\;\tilde{\underline a}\;|| = \beta \; ||\;{\underline a}\;||$, where 
$\alpha > 1$ and $\beta < 1$. 

The exact magnitude of $\alpha$ depends on the distribution of the
contravariant coefficients and the shift $\Delta x$. In fact, we can always
make the shift large enough so that $\alpha \gg 1$. In the same way,
we can ensure that $\beta \ll 1$. Assuming that the metric matrix $G$
has a minimum eigenvalue $e_{\rm min}$ and a maximum eigenvalue
$e_{\rm max}$, we then find:
\begin{equation}\label{MinMaxEigen}
e_{\rm min} \le \frac{||\;\tilde{\underline a}\;||}{||\;\tilde{\bar a}\;||}
            = \frac{\beta}{\alpha} \frac{||\;\underline a\;||}{||\;\bar a\;||} 
            \ll \frac{||\;\underline a\;||}{||\;\bar a\;||} 
            \le e_{\rm max}\ .
\end{equation}
Since $G$ is an infinite dimensional matrix there need not be a
maximum or a minimum eigenvalue. For the model problem introduced in
Sec.\,\ref{Induced}, we will give numerical evidence, in
Sec.\,\ref{CondNumb}, that the ratio of maximum and minimum
eigenvalues diverges. The $G$ matrix therefore is in some sense
singular.

\begin{figure}
\hspace*{30pt}\epsfxsize=240pt \epsfbox{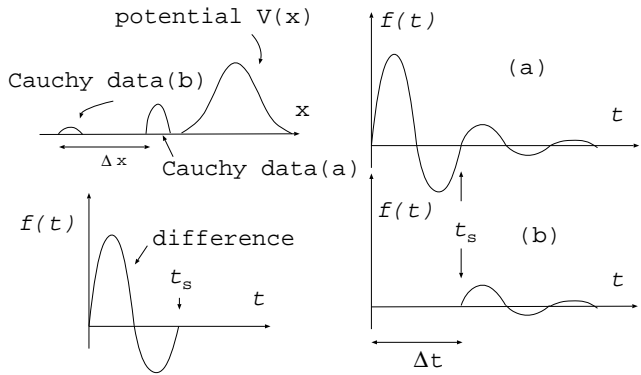}
\caption{\label{fig:shift2}
Oscillations produced by shifting and scaling initial data.}
\end{figure}

We now make the additional assumption that we have constructed
Cauchy data such that the wave form consists almost exclusively of
ringing in a single QN mode. (In the specific example discussed 
in the following sections,, this will be possible for the fundamental mode
using an arbitrarily short ``burst'' of initial data.)
Let us modify the
example of Fig.~\ref{fig:shift}, by reducing the size of Cauchy data
$(b)$ by a factor $\exp{(\Im{\omega_1})\Delta x}$, so that once the
response to the $(b)$ data starts it is identical to that of the $(a)$
response at the same time. This situation is shown in
Fig.~\ref{fig:shift2} in which the response $f(t)$ is the same in both
plots for $t\geq t_s=\Delta t$.
Now let us suppose that the waveforms are dominated by ringing of the
fundamental mode, and that $f(t)$ consists only of ringing of the
fundamental mode.  The difference between the two responses is a sum
of QN modes which has the appearance of truncated fundamental mode
ringing, but which has $a^1=a^{\tilde{1}} =0$ in its mode sum. The
time $t_s$ can in principle be made arbitrarily long, so the
truncation in the difference curve can be made arbitrarily late.  We
have then a sum of modes that looks arbitrarily close to fundamental
mode ringing, but containing no fundamental mode.  This suggests that
the fundamental mode can ``almost'' be built as a superposition of the
modes other than the fundamental mode, and that the modes are
``almost'' linearly dependent.  We will discuss this further, based on
numerical results, in Sec.\,\ref{BasAng}.

Again, this suggests that the infinite metric matrix $G_{nk}$ is in some
sense ``almost'' singular. 
A physical insight can be associated with this nature of the metric:
Since $G_{nk}$ is ``almost singular'' it is ``almost noninvertible.''
This means that there are difficulties in finding the contravariant
coefficients from the covariant coefficients, since this requires the
inverse of $G_{nk}$. The implication of this is possibly of pragmatic
importance: the near singularity makes it difficult to find the
``theoretical'' coefficients from the ``experimental'' ones.



\renewcommand{\RHd}{Quantifying QN excitations}  \markboth{\RHd}{\RHd}

\section {Induced Completeness}\label{Induced}

\subsection {Small change}

Our approach to quantification of excitation is based on the idea of
complete QN systems. It is important to demonstrate that complete QN
systems exist and to have an example of a system in which we can
compute excitation coefficients.  We must also deal with a more
specific issue.  The QN modes of potentials for black hole problems do
not form a complete set; both the Regge-Wheeler potential, which
describes perturbations of a Schwarzschild black hole, and the
P\"oschl-Teller potential\cite{PoschTeller}, which has been used as an
approximation to the black hole potential\cite{ferrarimashhoonPT},
have quasinormal mode sets which cannot completely describe the wave
form resulting from an initial perturbation. An important question is
what the relevance is to black hole processes, of any quantification
based on an assumption of complete systems. The details of this section
are somewhat disjoint from the main goal of the paper: quantification
of QN excitation, but are a necessary step in developing our approach.
(They are also rather interesting in their own right!)

The key idea in developing a demonstrably complete QN system, and in
showing its relevance to black holes, is ``induced completeness.''  It
has been argued \cite{youngsuenetal} that for the problem defined by
(\ref{eigeneq}), the eigenmodes will be complete if there are two
values of $x$ at which $V(x)$ is not $C^\infty$.  If this is true, it
appears that the black hole potentials can be modified in such a way
that completeness of QN oscillations is induced, while the effect on
other aspects of the problem -- in particular on the evolution of
Cauchy data -- is kept arbitrarily small.  In this section we explore
induced completeness.

It is not our purpose here to look into the generality of induced
completeness. Rather, we want to have a specific example of a complete
QN system with which to explore the question of quantifiability of QN
excitations. We will therefore focus on a very specific model that
contains most of the flavor of black hole potentials, but is simple
enough to allow rather straightforward analysis and a proof of
completeness.

We use the fact that (\ref{eigeneq})
can be solved with elementary functions in the case that the 
potential has the form of a truncated centrifugal potential,
\begin{equation}\label{truncpot}
 V_{\rm TCP}(x) = \cases{ 0 & $x < x_0$ \cr
                  \ell (\ell+1) / x^2 & $x \ge x_0$ \cr} ,
\end{equation}
where $\ell$ is an integer.  It gives a good representation of the
sharp cutoff of the black hole potential as $x\rightarrow-\infty$ and
the approximate asymptotic form of the black hole potential at
$x\rightarrow\infty$. It differs from the true black hole potential in
the details of the $x\rightarrow\infty$ potential that give rise to
the power-law late time tails of black hole waveforms. These tails
almost certainly are an obstacle to an analysis of QN excitation, and
a modification of the potential to eliminate these tails has already
been made in work on QN excitation.

We will hereafter consider only the case $\ell=1$, to be called the
``truncated dipole potential'' (TDP).  An example of the simplicity of
the TDP problem is that it has only a single pair of QN frequencies:
$\omega_1=(1+i)/2x_0$ and $\omega_{\tilde{1}}=(-1+i)/2x_0$ 
(see Appendix IA). 
There are
several ways in which we could try to induce completeness into this
problem. We have studied both the addition of a small step (with
discontinuities) to the TDP and the addition of a delta function.  The
delta function has the disadvantage of its distributional nature, but
it offers the advantage of considerable simplicity as compared with the
step. We have found no significant ``practical'' difference between the 
results (QN locations, convergence of QN sums) between the two examples, so
we will describe here completeness induced with a delta function. The total
potential in this case will be called the ``spiked truncated dipole
potential'' (STDP) and is given by
\begin{equation}
  \label{stdpdef}
  V_{\rm STDP}(x) = V_{\rm TDP}(x) 
                      \big( 1 + V_\delta\,\delta(x-x_\delta) \big) .
\end{equation}
where 
$V_{\rm TDP}(x)$ is the $\ell=1$ form of the potential in (\ref{truncpot}).

We first establish that the influence on the evolution of initial data
can be made arbitrarily small. To do this we choose the Cauchy data, at $t=0$
to be given by
\begin{eqnarray}\label{inidat}
\psi_0 &=& \sin{\left(2\pi {x-x_2 \over x_1-x_2}\right)} \quad 
               (x_2 \le x' \le x_1) \\
\dot{\psi}_0
&=&
-\partial\psi_0/\partial x\nonumber\ ,
\end{eqnarray}
with $x_1\leq x_0$, so that the initial data has the form of one full
cycle of a sine wave, located to the left of the potential, traveling
to the right. (This Cauchy data is chosen for convenience in demonstrating
mathematical points; it has no justification as initial data for gravitational
waves being produced in the neighborhood of a black hole.)

For the results to be shown here and in following sections, we choose
$x_0=1$,  $x_1=1$, $x_2=-5$, and $x_\delta=10$, and $x_{\rm obs}=120$.
For this choice of $x_1$ and $x_2$, the sine has a wavelength, 6, that is 
roughly half that of the QN oscillation of the \TDP\ . This allows us to 
see similar, but distinguishable signs in the evolved waveform of the 
propagation of the Cauchy data and of QN oscillation. 

\begin{figure}[bth]
  \hbox{\epsfxsize=0.49\hsize\epsfbox{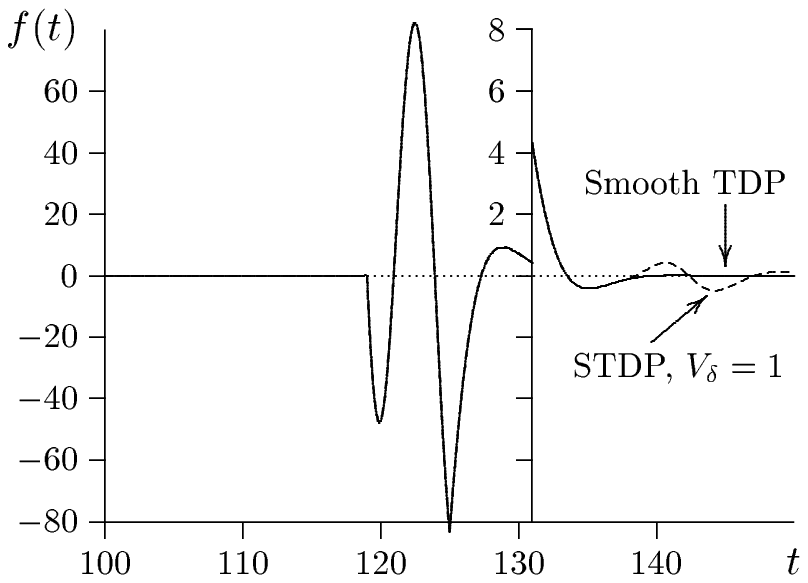} \hfil
        \epsfxsize=0.49\hsize\epsfbox{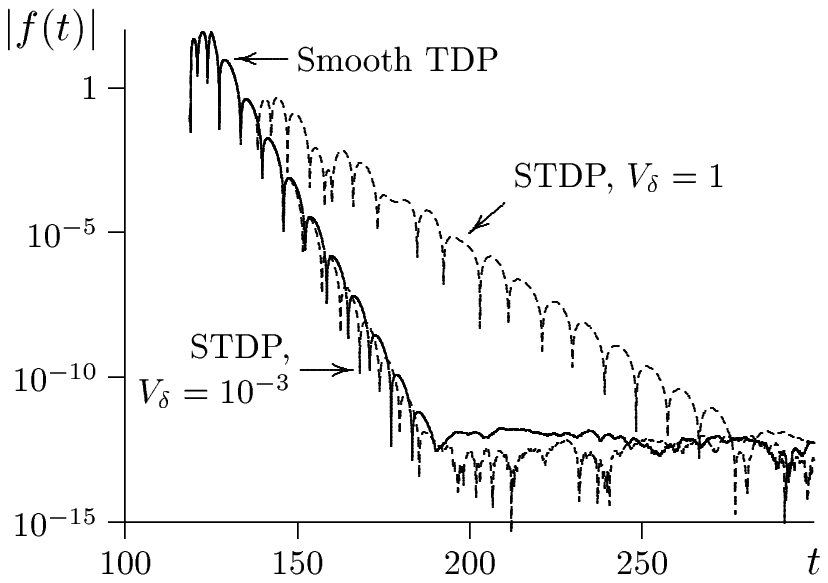}}
\vspace{0.5\baselineskip}
\caption[Text to appear in the list of figures] 
{
Time evolution for initial data incident on the \TDP\ and on the \TDP\ 
with an added $\delta$-function. Left: linear plot, right: logarithmic
plot}
\label{deltim}
\end{figure}

Figure \ref{deltim} shows the time evolution of the Cauchy data for
the original \TDP\ and for the \TDP\ with an added $\delta$-function
with different amplitudes ranging from $V_\delta =1$ to $V_\delta
=10^{-6}$.  The waveforms are followed out to times at which they have
decreased in magnitude from the maximum by a factor of $\sim
10^{-15}$, at which point numerical error obscures the results.  Even
for the largest value of $V_\delta$, the influence of the delta
function on evolution is visible in the linear plot only after we
change the scale for $f(t)$ at $t = 131$. To see the effects more
clearly we also plot the logarithm of $|f(t)|$ for a longer
observation time, showing several interesting features. For all but
the largest amplitude of the delta function ($V_\delta=1$) the
waveform after the first few oscillations consists only of QN ringing,
with the characteristic distance $\Delta t= |\Re{\omega_7}|\pi=2\pi
x_0=2\pi$ between zeroes of $f(t)$. For $V_\delta=10^{-3}$, the effect
of the delta function shows up only as a phase shift after the the
first four, or so, full cycles of oscillation. The effect of the
smallest delta function amplitude $V_\delta=10^{-6}$ is smaller by two
orders of magnitude than that for $V_\delta=10^{-3}$, and too small to
be seen even in the logarithmic plot of Fig.\,\ref{deltim}.

These results make it clear that any reasonable measure of the
influence of the delta function, such as the integral of the square
deviation from the TDP waveform, is tiny and decreases with decreasing
$V_\delta$.  We will accept these numerical results as a sufficient
demonstration of this point, and will not attempt an analytic proof of
this point.

\subsection {QN spectra}

Although the influence of a small delta function on the evolution of
Cauchy data is negligible, the influence on the spectrum of QN
frequencies is profound. The method of computing the QN frequencies
for the STDP is outlined in Appendix I. The results of the
computation are presented in Fig.\,\ref{delqnm} for the same
potentials considered in Fig.\,\ref{deltim}.

The original \TDP\ has only one pair of QN frequencies at
$(\pm1+i)/2$.  With an added $\delta$-function, an entirely new set of
modes appear. Note that the real parts of the additional frequencies
seem to be unbounded. The asymptotic spacing $\Re{(\omega_n)}=2\pi/10$
of the real parts of the frequencies is shown in 
Appendix IC to correspond to the length of the ``cavity''
bounded by $x_0$ and the $\delta$-function: $L\equiv x_\delta-x_0=10$.
The imaginary parts are shown in the Appendix to increase as the
logarithm of the real part.

When $V_\delta$ becomes sufficiently small, one of the QN frequencies
approaches the value $(1+i)/2x_0$ of the ``native'' QN frequency, that
of the original, \TDP\ . This can be seen more clearly in Fig.
\ref{delqnm1}, which shows the ``path'' of this QN frequency in the
complex plane for values of $V_\delta$ varying in steps of $1/2$
from $V_\delta=1$ to $V_\delta \sim 10^{-6}$.  As $V_\delta$ decreases, the
imaginary parts of the additional frequencies increase, moving them
away from the native one, eventually leading to two very distinct
subsets of QN frequencies.

\begin{minipage}[bth]{0.98\hsize}
\begin{minipage}[t]{0.49\hsize}
\begin{figure}[bth]
   \epsfxsize=\hsize\epsfbox{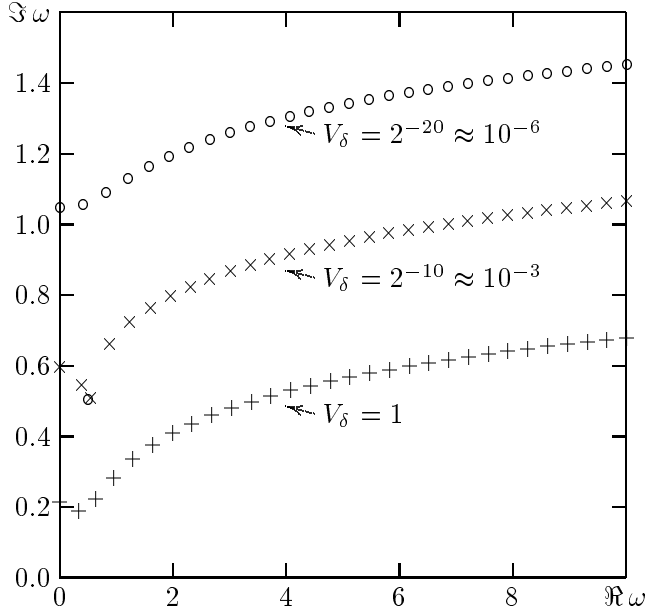}
\vspace{0.5\baselineskip}
\caption[Text to appear in the list of figures] 
{
Quasinormal frequencies for the \TDP\ with an additional
$\delta$-function at $x_\delta = 10$.
}
\label{delqnm}
\end{figure}
\end{minipage}
\hfil
\begin{minipage}[t]{0.49\hsize}
\begin{figure}[bth]
   \epsfxsize=\hsize\epsfbox{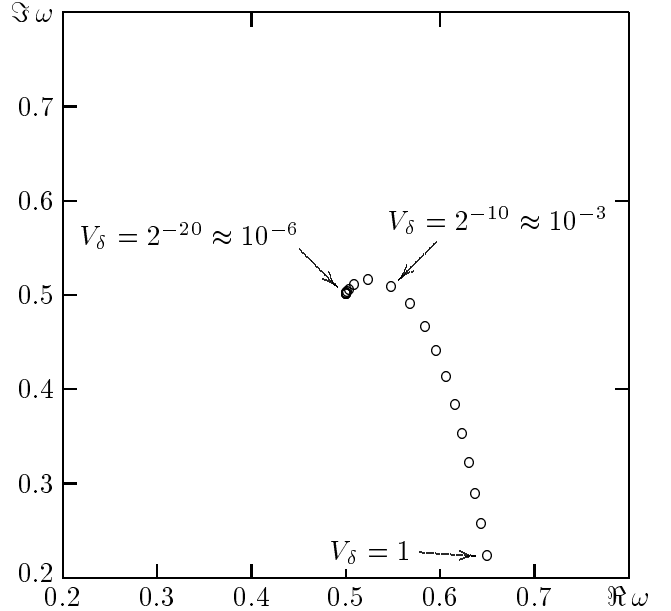}
\vspace{0.5\baselineskip}
\caption[Text to appear in the list of figures]
{
``Path'' of the original quasinormal frequency of the \TDP\ as the
amplitude of the additional $\delta$-function is decreased.
}
\label{delqnm1}
\end{figure}
\end{minipage}
\vspace{\baselineskip}
\end{minipage}

It might seem that it is an obvious necessity for the QN spectrum to
have a mode approximately at the location of the native mode, since
the evolution of Cauchy data is only slightly affected. This turns out
not to be true, however, for other ways of inducing apparent
completeness\cite{StepPot}.  Cutting off the TDP potential at some
very large value of $x$, for example, has a negligible effect on the
evolution of Cauchy data, and it also produces an unbounded set of
additional QN frequencies. However, the spectrum of QN frequencies
turns out to contain no frequency near the location of the native
mode.

\subsection {Complete sums}

We now develop  the connection between the Cauchy data $\psi_0$ and
$\dot{\psi}_0$ and the coefficients of QN oscillations (i.e., the
contravariant components $a^n$ in the case of complete QN bases).  
Here we simply outline how coefficients associated with  QN
basis functions are computed in general.
In
Appendix II a proof of completeness of sums with these coefficients
will be given for the case of the spiked TDP, and of Cauchy data meeting
certain criteria. One of the criteria will be compact support for 
the Cauchy data, and we will assume from the outset that $\psi_0$ and
$\dot{\psi}_0$ vanish outside the interval $x_2<x<x_1$.

The QN coefficients are found by starting with a Laplace transform
\begin{equation}
\Lpsi(s,x) = \int_0^\infty e^{-st}\ \Psi(t,x)dt \ .
\end{equation}
(The connection between the Laplace variable $s$ and the Fourier $\omega$
used in Appendix I and  most of the paper, is through the correspondence
$s\leftrightarrow i\omega$.)
The transformed wave equation reads
\begin{equation}\label{Laplwave}
\partial\Lpsi(s,x)/\partial x^2 -\left[s^2+V(x)\right]\Lpsi(s,x)=J(s,x)\ ,
\end{equation}
in which the source  $J(s,x)$ is determined by the Cauchy data:
\begin{equation}
J(s,x) = - \dot\psi_0(x) - s\psi_0(x)
\end{equation}
A solution can be found in the form
\begin{equation}\label{GFsoln}
\Lpsi(s,x) = \int_{x_2}^{x_1} G(s,x,x') J(s,x') dx'\ .
\end{equation}
Here the Green function can be constructed from the homogeneous
solutions $f_-(s,x)$ and $f_-(s,x)$ of of the wave equation
(\ref{Laplwave}):
\begin{equation}\label{GFconstruction}
G(s, x,x') = {1 \over W(s)}
                    \cases{ {f_-(s,x') f_+(s,x) \quad (x' < x) , } \cr
                      \noalign{\smallskip}%
                            {f_-(s,x) f_+(s,x') \quad (x' > x) , } \cr }
\end{equation}
where $W(s)$, the Wronskian of $f_-$ and $f_+$ is independent
of $x$ and $x'$.

We will assume that the potential falls off quickly enough at
$x\rightarrow\pm\infty$ so that homogeneous solutions $f_-$ and $f_+$
can be found with the property
\begin{equation}\label{boundconds}
 f_-(s,x) \sim e^{sx}\left(1+{\cal O}(1/|x|)\right) \quad \hbox{as} \quad x \to -\infty
           \qquad  \hbox{and}  \qquad
   f_+(s,x) \sim e^{-sx}\left(1+{\cal O}(1/|x|)\right) \quad \hbox{as} \quad x \to \infty\ .
\end{equation}
This condition is satisfied for black hole potentials and for the TDP.

Once we have found a solution of the Laplace transformed wave
equation, we can reconstruct a solution of the time-dependent wave
equation by applying the inverse Laplace transform: 
\begin{equation}
\Psi(t,x) = { 1 \over 2\pi i} \int_\Gamma e^{st} \Lpsi(s,x) ds
\end{equation}
where $\Gamma$ denotes the path of integration, which lies parallel to and
just to the right of the imaginary axis.

\begin{figure}[bth]
  \hspace*{1.2in} \epsfxsize=.4\hsize\epsfbox{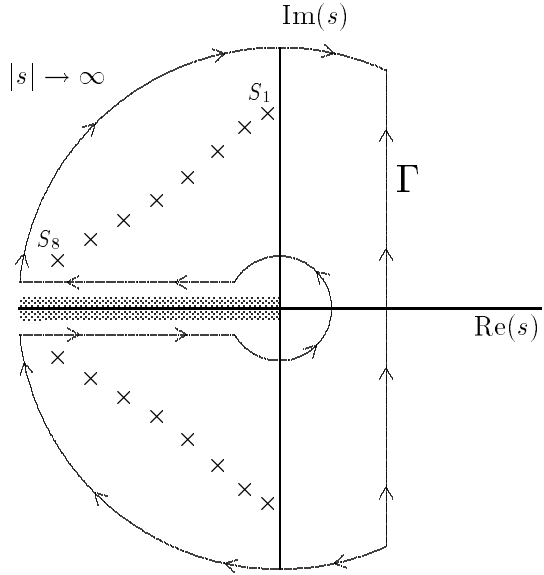}
\vspace{0.5\baselineskip}
\caption[Text to appear in the list of figures] 
{
  Closing the contour for integration in the complex $s$ plane. The
  curve $\Gamma$ is the original contour of integration. It can be closed
by the addition of an arc at infinity. If there is a branch point at the 
origin, as shown, then a branch cut can be drawn from $s=0$ to
$s=-\infty$, and the contour shown can be drawn.}
\label{intpath}
\end{figure}
Figure\,\ref{intpath} illustrates features in the complex $s$ plane.
There may be poles and essential singularities, drawn as $\times$s, and
there may be a branch point, like that shown at $s=0$ in the figure.
The contour $\Gamma$ can be closed with the addition of a single arc to the
left if $\Lpsi(s,x)$ is a single valued function of $s$ to the left of
$\Gamma$.  If there are branch points then a path like that shown in 
Fig.\,\ref{intpath} must be used. 
The Green's function for the Regge-Wheeler potential has a branch
point at $s=0$, while the TDP and the STDP do not.

The integral for $\Lpsi$ may then be evaluated as 
\begin{equation}\label{sum_int}
\Psi(t,x) = \sum {\rm Res}(e^{st} \Lpsi(s,x),s_j)
           + { 1 \over 2\pi i} \int_{\cal C} e^{st} \Lpsi(s,x) ds
\end{equation}
where the first term is the sum of residues at the singularities
inside the completed contour. The second term represents contributions
to the integral from arcs at $\infty$, and along branch cuts. In the
case of the spiked TDP there is no branch cut, and the arc at $\infty$
makes no contributions, so we are left only with the first term, a sum
of oscillations at discrete frequencies $s_j$ that correspond to
singularities of $\Lpsi(s,x)$. From (\ref{GFsoln}) and
(\ref{GFconstruction}) we see that singularities in $\Lpsi(s,x)$ must
either be singularities of the homogeneous solutions $f_+,f_-$ or
zeroes of the Wronskian. For any finite $s$ we can find homogeneous
solutions $f_+$ and $f_-$ of (\ref{Laplwave}); therefore,
singularities in the Green's function can occur only at zeros of
$W(s)$, and the residues in (\ref{sum_int}) must be taken at these
zeros. But the vanishing of $W(s)$ at $s_j$ means that $f_+\propto
f_-$ at $s_j$, and hence that $s_j$ is a QN
frequency~\cite{NollertSchmidt}.

If, as in the case of the sTDP, the only contributions to
(\ref{sum_int}) occur in the sum, then we are left with
\begin{equation}\label{sumofqnf}
\Psi(t,x) = \sum {\rm Res}(e^{st} \Lpsi(s,x),s_j)\ ,
\end{equation}
where the sum is over QN frequencies. 

We now assume that the zeros of $W(s)$ are all first order, so that 
\begin{equation}
  \label{resatqn}
  W(s)=\left.\frac{dW}{ds}\right|_{s=s_j}\hspace*{-8pt}(s-s_j)
+{\cal O}[(s-s_j)^2]\ .
\end{equation}
For $x>x_1$ (i.e., for $x$ to the right of the region in which the 
Cauchy data have support) we  have,
from (\ref{GFconstruction})
 that 
\begin{equation}
  \label{resofG}
  G(s,x,x')=\frac{f_-(s_j,x')f_+(s_j,x)
}{dW/ds|_{s_j}(s-s_j)}+{\cal O}\left(s-s_j\right)^0
\end{equation}
and hence, from 
(\ref{GFsoln}), the residues of $\Lpsi$ at $s_j$ is
\begin{equation}
  \label{resofpsi}
{\rm Res}(\Lpsi(s,x) =\frac{e^{s_jt}f_+(s_j,x)
}{dW/ds|_{s_j}
}
\int_{x_2}^{x_1}J(s_j,x')f_-(s_j,x')\,dx'\ .
\end{equation}
Finally, the sum over the QN basis functions can be written
\begin{equation}
  \label{finalsum}
  \Psi(t,x)=\sum b^j u_j(t,x)\ ,
\end{equation}
where 
\begin{equation}
u_j(t,x) = f_+(s_j,x) e^{s_j t}
\end{equation}
and
\begin{equation}\label{coeff_formula}
b^j = -\frac{1}{\left. dW/ds\right|_{(s_j)}
}
      \int \left(\dot\psi_0(x') + s_j \psi_0(x')\right) f_-(s_j,x') \, dx'\ .
\end{equation}
The waveform at $x_{obs}>x_1$ is then given by (\ref{compdef})
with 
\begin{equation}
  \label{notationchange}
  i\omega_j=s_j\quad a^j=b^jf_+(s_j,x_{obs})\ .
\end{equation}

This prescription has been applied to the spiked TDP, and the sine wave
Cauchy data, with our standard choice of parameter values, $x_0=1$,
$x_\delta=10$. 
Notice that our choice $x_2=-5$ in (\ref{inidat})
satisfies the criterion in the convergence proof that the continuous,
but nonsmooth, Cauchy data have support only for $x > x_0 -
2/3(x_\delta-x_0)=-5$.  For this potential and these Cauchy data, the
functions $\psi_0(x)$,$\dot{\psi}_0(x)$ and $f_-(s_j,x)$ are
trigonometric or exponential functions, so the integral in
(\ref{coeff_formula}) is elementary.  Once the QN frequencies,
$\omega_j$, and the factors $dW/ds|_{\omega_j}$ have been computed (as
described in Appendix I), the coefficients are easily evaluated. The
figures below show the result of using these coefficients in sums of
the form (\ref{compdef}).

Figure\,\ref{ms_E0} shows the computed result for time evolution of
the Cauchy data in the case that $V_\delta = 1$. This is compared
along with mode sums for an increasing number of terms, up to
$N=10,001$ terms. Figure\,\ref{ms_E6} shows the same plots in the case
that $V_\delta = 10^{-6}$. In order to avoid cluttering the picture,
we do not plot the values of the mode sums when they are far from
having converged.
These figures illustrate the fact that  the mode sum converge
faster for a larger $\delta$-function. Also, for the smaller
$\delta$-function, the mode sum  converge more rapidly at later
times, with convergence ``sweeping down'' from late to early
times.

\begin{minipage}{0.98\hsize}
\begin{minipage}[t]{0.49\hsize}
\begin{figure}[bth]                 
   \epsfxsize=\hsize\epsfbox{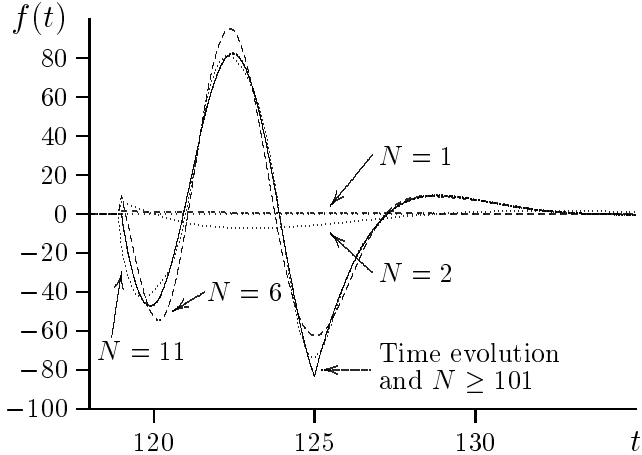}
\vspace{0.5\baselineskip}
\caption[Text to appear in the list of figures]
{Values of the mode sum for different number of terms $N$, compared to the
wave form resulting from integrating the time dependent wave
equation ($V_\delta = 1$).} 
\label{ms_E0}
\end{figure}
\end{minipage}
\hfil
\begin{minipage}[t]{0.49\hsize}
\begin{figure}[bth]                  
   \epsfxsize=\hsize\epsfbox{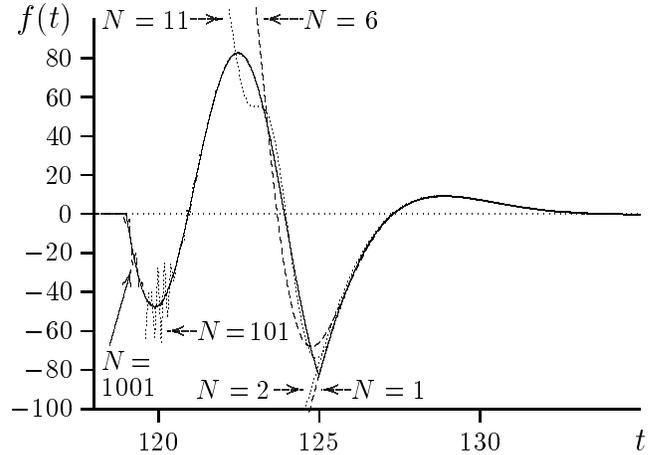}
\vspace{0.5\baselineskip}
\caption[Text to appear in the list of figures]
{Values of the mode sum for different number of terms $N$, compared to the
wave form resulting from integrating the time dependent wave
equation ($V_\delta = 10^{-6}$).} 
\label{ms_E6}
\end{figure}
\end{minipage}
\vspace{\baselineskip}
\end{minipage}

For a better view of the differences between the mode sums and the
time evolved wave functions, Fig.\,\ref{diff_ms_E6} shows the
logarithmic difference (evolved waveform vs. mode sum) for $V_\delta =
10^{-6}$. 

Some of the systematics shown in these figures can be
heuristically understood. The convergence in the $V_\delta = 1$ case
is similar to that of a fourier series (equally rapid at different
times) since the QN frequencies, in this case, with small imaginary
parts, are similar to the real frequencies of a fourier series.  On
the other hand, for $V_\delta = 10^{-6}$, the damping of the
additional (i.e., non-native) modes is much stronger and increases
faster with $N$. Any error that originates from cutting off the mode
sum after a finite number of terms can be regarded as being composed
of modes with very strong damping; these modes are very large at early
time.

We point out here an interesting technical feature of these results.
The mode sums are so accurate at later times that the differences
shown in Fig.\,\ref{diff_ms_E6} for $N=10,001$, is actually dominated,
after $t\approx 122$, by the numerical 
truncation error in computing the evolved waveform.
A smaller time step in numerical evolution can improve the numerical
accuracy of the computed waveform, and can move to a slightly larger time
the point at which the evolution vs. sum difference is dominated by
the numerical errors in the evolution.

\centerline{
\begin{minipage}{0.49\hsize}
\begin{figure}[bth]             
  \epsfxsize=\hsize\epsfbox{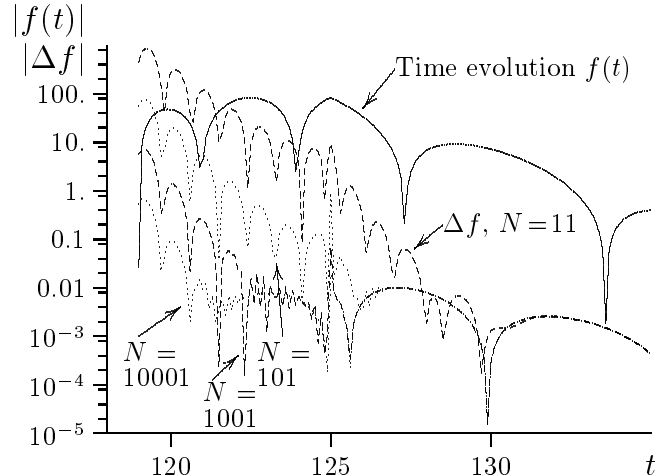}
\vspace{0.5\baselineskip}
\caption[Text to appear in the list of figures]
{Logarithmic difference $\Delta f$ between the values of the mode sum for
different numbers of terms and the 
wave form resulting from integrating the time dependent wave
equation ($V_\delta = 10^{-6}$).} 
\label{diff_ms_E6}
\end{figure}
\end{minipage}
}
\medskip

%

\renewcommand{\RHd}{Quantifying QN excitations}  \markboth{\RHd}{\RHd}

\section{Quantifying excitation}\label{Quantify}

\subsection {The excitation coefficient}

We now return to the question of how to quantify the excitation of QN
oscillation. In Sec.\,II we defined the excitation coefficient $A_k$
for a QN oscillation in a complete QN system. Due to the time shift
problem, we argued that the excitation coefficient seems the only
plausible indicator of the QN content of a waveform. In Sec.\,III we
have seen that, at least in the particular example of the \TDP,
completeness can be induced in an incomplete system to create one that
is ``physically equivalent,'' i.e., differs negligibly in the
evolution of Cauchy data. We can now ask whether, at least for the
model problem at hand, we can use the excitation coefficients of the
completed system to quantify the QN oscillation in the original
(``native'') system.

It is worth emphasizing that the \TDP\ is a particularly simple
starting point for these considerations, since it has only a single
conjugate pair of native QN modes. The sine wave data we have used in
the previous sections is also convenient since it produces a waveform
(see Fig.\,\ref{deltim}) which clearly contains QN
ringing, but contains significant oscillation at a different frequency
($\omega=2\pi/6$).

The results of computations with this model are
presented in Table\,\ref{tabcoeff}. Results are shown for the \sTDP\
for different values of $V_\delta$ and for the standard Cauchy
data, a right moving sine wave initially extending over the interval
$[-5,1]$.  For comparison, a shifted sine wave, initially at
$[-8,-2]$, was also computed in the case $V_\delta=10^{-6}$. 
Also included are results for the smooth TDP, and for the Zerilli
equation with the Regge-Wheeler potential with initial data
corresponding the the Close Limit technique for black hole
collisions~\cite{prpu}. 

The table presents values of excitation coefficients
$A_1\equiv a^1(a_1)^*+(a^1)^*(a_1)$ and the energy excitation
coefficients 
$E_1\equiv(\omega_1^*)^2(a_1)^*a^1
+(\omega_1)^2(a_{\tilde{1}})^*a^{\tilde{1}}$, 
introduced in (\ref{excoeffdef}) and
(\ref{defenexcoeff}). In each case the QN frequency $\omega_1$ is
taken to be that eigenfrequency which corresponds to the native mode
in the limit $V_\delta\rightarrow0$; that is, $\omega_1$ always lies on
the path in the complex plane shown in Fig.\,\ref{delqnm1}. For small
values of $V_\delta$ this QN frequency is close to the native QN
frequency $(\pm1+i)/2$.  The waveform norm and the total energy are
computed from wave forms obtained by explicit numerical integration of
the wave equation (\ref{basiceq}). Also included are estimates for the
numerical uncertainties of the results.

The norm and total energy can be computed from Eq.(\ref{sumofsq}), or
directly from the wave form. Similarly, the covariant coefficient may
be computed from the contravariant ones, using Eq.(\ref{covfromG}), or
from the wave form via Eq.(\ref{covdef}). Even the contravariant
coefficients can be obtained from the wave form itself, using an
asymptotic fit as it approaches a pure QN oscillation at the least
damped QN frequency, instead of the residues of the Green's function
through Eqs.~(\ref{coeff_formula}) and (\ref{notationchange}).  The
wave form itself, in turn, can be obtained by direct numerical
integration of the wave equation, or by using the mode sum as in
Eq.(\ref{compdef}). We have checked these alternatives for obtaining
the quantities listed in the table; they result in essentially the
same values as the route we have taken here.

We first notice that the values for the excitation coefficients are
identical, within the numerical errors, for the original initial data
(sine wave over the interval $[-5,1]$) and the shifted initial data
(over the interval $[-8,-2]$). This confirms our earlier argument that
the excitation coefficients we have defined are independent of a
translation of the initial data or, correspondingly, to a time shift
of the evolved wave form.

%
%
\begin{table}[tbh]
\begin{center}
\begin{tabular}{|l|l|l|l|l|c|c|}
$V(x)$ & $V_\delta$ & Initial data& 
  ~$A_1$~ & ~$E_1$~ & ~$\bigl| f \bigr|^2$~ &
  ~$\bigl| \dot f \bigr|^2$~ \\
\tableline
%
%
STDP & $10^{-3}$ & Sine $[-5,1]$  & 
       $- 40\;854. \pm 1. $ & $ - 23\;810. \pm 1. $ & 
       $  19\;246. \pm 7. $ & $  27\;404. \pm 12. $ \\
STDP & $10^{-6}$ & Sine $[-5,1]$ &  
       $- 57\;933. \pm 1. $ & $ 5\;804.7 \pm 0.1 $ & 
       $  19\;246. \pm 7. $ & $ 27\;404. \pm 12. $ \\
STDP & $10^{-6}$ & Sine $[-8,-2]$ &
       $- 57\;934. \pm 1. $ & $ 5\;804.5 \pm 0.3 $ & 
       $  19\;246. \pm 7. $ & $ 27\;404. \pm 12. $ \\
STDP & $10^{-7}$ & Sine $[-5,1]$ &  
       $- 57\;879. \pm 1. $ & $ 5\;811.9 \pm 0.1 $ &           &           \\
STDP & $10^{-8}$ & Sine $[-5,1]$ &  
       $- 57\;873. \pm 1. $ & $ 5\;812.6 \pm 0.1 $ &           &           \\
STDP & $10^{-9}$ & Sine $[-5,1]$ &  
       $- 57\;873. \pm 1. $ & $ 5\;812.6 \pm 0.1 $ &           &           \\
TDP &  --        & Sine $[-5,1]$ &  
       $- 57\;872.8\ldots $        &   $5\;812.64\ldots$      & $19\;248.0\ldots $    & 
       $  27\;415.5\ldots $ \\
\tableline
Zerilli  &  --        & Close Limit &  
                  $10.699 \times 10^{-4}$ &     & $5.566 \times 10^{-4}$ & \\
\end{tabular}
\end{center}
\caption{\label{tabcoeff} 
Excitation coefficients 
and norms for the \sTDP\ QN system and for black holes in the
close limit approximation.}
\end{table}


It is important to realize that
there are different ways to compute the quantities in the table, and
several of them do not depend on the complete set of modes. The
waveform and energy norm, of course, require only the waveform, but
all coefficients can also be computed entirely from the waveform
itself. The least damped QN frequency can be inferred from the
asymptotic late time behavior, as can the contravariant coefficient
for the least damped mode. [See (\ref{compdef}).]  Of the cases we
study the least damped mode always corresponds to the native mode
except for the $V_\delta=1$ model. (See Fig.\,\ref{delqnm}.)  Once the
QN frequency is known the covariant coefficient can be computed
directly from the integral in (\ref{covdef}) and, with the
contravariant coefficient known, the excitation and energy
coefficients can then be found from (\ref{excoeffdef}) and
(\ref{defenexcoeff}).  Since the quantities in the table can be
computed from waveforms, we can compute them for the smooth TDP. In
this case we do not have a complete set of modes, but that is
irrelevant to the procedure for computation.  (It turns out, in fact,
that the simplicity of the smooth TDP and the sine wave data allows a
closed form solution for the waveform, and for the norms and
coefficients.  This closed form solution has been used, and the values
for the smooth TDP can be found for the table to essentially arbitrary
precision.)

The numbers in the table make it clear that the results for the smooth
TDP are the $V_\delta\rightarrow0$ limit of the \sTDP. This is obvious
from a computational point of view, since all results can be computed
from the waveform, and the $V_\delta\rightarrow0$ waveform approaches
the smooth TDP waveform. From another point of view, however, this
result is interesting and important. It means that we can compute the
excitation coefficient independent of the method with which completeness
is induced. Put another way, it means that we can compute the
excitation coefficient for a small completeness-inducing change,
independent of the nature of the change. This conclusion, in fact, is
crucial to the possible importance of the excitation coefficient. We
could not use the excitation coefficient to characterize an excitation
of a physical system if the value of the coefficient depended on our
choice of a modification of that physical system.

The table also gives a value for the
excitation coefficient for the gravitational radiation produced, in the
close limit approximation\cite{prpu}, by the head on collision of two
equal mass nonrotating holes. In this case the QN spectrum is
infinite, but the QN oscillations are not complete. The excitation
coefficient given is for the least damped of the QN oscillations, a
frequency that appears to dominate the waveform produced.

Up to this point we have noted that computations with the model
problem illustrate and confirm the features of the excitation
coefficient that made it an attractive candidate for the
quantification of the excitation of QN oscillations.
Table\,\ref{tabcoeff}, however, also makes it unlikely that the
excitation is a {\em useful} index of QN oscillation. For the
small-$\delta$ \sTDP\ the excitation coefficient is negative, and is larger
(by a factor $\sim 3$) in magnitude than the norm. Note that we cannot
ignore this ``wrong'' sign, and simply keep the large magnitude as an
indication of a strong QN presence. Due to relation (\ref{sumofsq}) we
must conclude that the sum of all the other QN oscillations (those
unrelated to the native QN mode) must be greater (by a factor $\sim 4$)
than the norm. For the close limit waveform the excitation coefficient
is roughly twice the size of the norm of $f$ and hence the sum of
excitation coefficients of all other QN oscillations (if the system
were made complete somehow) would be negative.

As a possible alternative to the excitation coefficient we have also
computed the energy excitation coefficient, as defined in
(\ref{defenexcoeff}). The results listed for $E_1$ and the norm of
$\dot{f}$, however, do not make this any more attractive as a measure
of QN excitation. 

At this point, we also note that the excitation coefficient $A_1$ is
not related to the norm of the QN mode contribution corresponding to
the first pair of modes, i.e. of $ b^1 u_1 + (b^1 u_1)^* $. 
Similarly, the energy excitation coefficient $E_1$ is no measure
for the energy of this QN mode contribution.

It is, of course, impossible to prove that something like quantifying
QN excitations cannot be done in a mathematically acceptable way.
Despite this, and perhaps to provoke further work (by others!) we are
tempted to offer the following very tentative conclusion: 
Consider the following three conditions, which allow a mathematically
meaningful as well as a useful measure of excitation; they are all
satisfied for normal mode systems:
\begin{itemize}
\item The measure of excitation is independent of a simple shift of
the wave form (i.e. a shift in space of the initial data,
corresponding to a shift in time for the time evolved wave form).
\item Excitation strength can be quantified individually for any
number of modes, with the individual excitations adding up to the
total norm of the wave form.
\item The measure of excitation is useful for quantifying the
excitation in a comparative way; in particular, it always lies between
0\% and 100\%.
\end{itemize}
We conjecture: There is no quantification of QN oscillations of the
waveform, based on the algebra of the function space of QN
oscillations, which satisfies all three of these criteria.

The excitation coefficients we are defining in this paper satisfy the
first two conditions, but not the third one (In Table\,\ref{tabcoeff}
we see that the excitation coefficient can be negative and can be 
greater than 100\%.)
One might regard these
two conditions as related to mathematical properties of the QN mode
system, while the third one is of a more practical significance. 
We
are currently investigating another technique to quantify excitations,
with a measure that satisfies the first and third
criteria, but not the second one. This measure may turn out 
to be of some practical utility, but is not as closely related to 
the mathematics of the function space as are the present considerations.
A description of this work will be published elsewhere.

Our conjecture is based specifically on the appearance of the
excitation coefficient as the only quantity in that function space
that solves the ``time shift'' problem, and on the observed failure of
the excitation coefficient to be ``useful.''  It is also based, less
specifically, on our belief that there are differences between normal
mode and QN systems that cannot be bridged. For this reason we
speculate that inducing completeness, while it is mathematically
interesting, is not likely to lead to useful tools for understanding
the underlying native system. The reason is that the spectrum of added
QN frequencies is unrelated to the native system and characterizes
only the method used to induce the completeness. We, in fact, are
willing to extrapolate in the direction of this speculation. We
conjecture that in a complete QN spectrum which has a dense set of
frequencies, and a small number of ``isolated'' frequencies, like the
spectrum of the \sTDP\ s in Fig.\,\ref{delqnm}, it is useful to modify
the system to {\em remove} completeness, in order to get a more useful
understanding and a simplified method of analysis. The obvious example
of this is removing the $\delta$ functions from the \sTDP\ problem in
order to get a physically equivalent, but simpler incomplete system.

A very similar point of view is that in a QN spectrum, not all
frequencies are equally important. Some will actually be evident in
waveforms produced in the evolution of generic Cauchy data; others
won't. For the small-$\delta$ spectra of Fig.\,\ref{delqnm}, the
``interesting'' QN frequencies are the isolated ones near $(1+i)/2$.
In the $V_\delta=1$ case, on the other hand, there is again the
appearance of QN ringing in the waveform, but the spectrum contains no
isolated frequency. A similar situation was found in a study where the
Regge-Wheeler potential was replaced by a series of step
potentials~\cite{StepPot}.  More generally, one should ask: If one has
only the spectrum and the associated quantities (e.g., the metric
matrix), is there a way of identifying which QN frequencies are
``important'' in the sense of really characterizing the evolution of
Cauchy data? In this sense we are asking a question related to ``to
what extent are (some) QN modes like normal modes?'' since normal
modes {\em do} characterize the system in which they arise.

\subsection{Condition number of the metric matrix}\label{CondNumb}

The metric matrix (\ref{metricdef}) would appear to be a likely place
to find a way of characterizing a QN system without regard to specific
waveforms produced by specific Cauchy data. We will now discuss some
numerical results which confirm the intuitive insight we had gained by
doing thought experiments on specific Cauchy data in Sec.~\ref{Intuit}. 
Studying the metric matrix directly, we don't need to refer to
specific initial data any more, as we had to do before.

We first note the singular
nature of the infinite metric matrix in two cases. To characterize
this singular property we truncate the set of QN functions, keeping
only the first $D$. We then compute the condition number $R$ (ratio of
maximum to minimum eigenvalue) for the $D$ dimensional subspace. The
condition number as a function of $D_{\rm sub}$ is plotted in
%
\centerline{
\begin{minipage}{0.49\hsize}
\begin{figure}[bth]
   \epsfxsize=\hsize\epsfbox{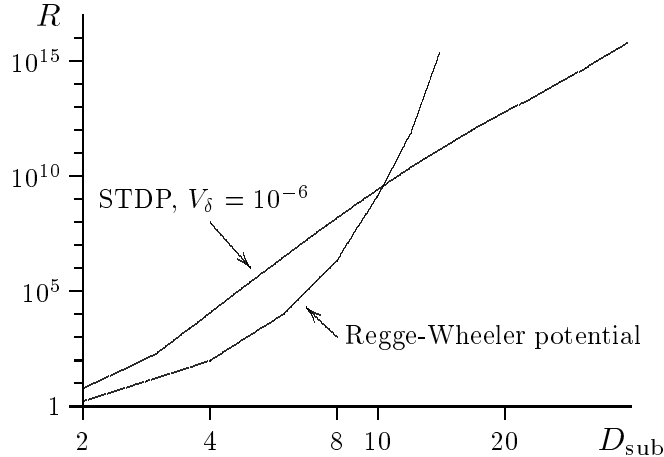}
\vspace{0.5\baselineskip}
\caption[Text to appear in the list of figures]
{The condition number $R$ for the metric matrix for a truncated subspace
of dimension $D_{\rm sub}$, spanned by the first $D_{\rm sub}$ QN
basis functions. Results are shown for both the \sTDP\ ($V_\delta =
10^{-6}$) and the Regge-Wheeler potential.}
\label{fig-cond}
\end{figure}
\medskip
\end{minipage}
}
%
Fig.\,\ref{fig-cond} for two cases: the \sTDP\ and the QN spectrum of
Schwarzschild black holes (the QN modes of the Zerilli or
Regge-Wheeler potentials). For the \sTDP\ the approximately straight
line in the log-log plot suggests that the condition number increases
roughly as the twelfth power of the dimension $D_{\rm sub}$ of the
subspace. For the black hole QN spectrum the increase is even more
dramatic, suggesting perhaps a ``more singular'' metric for this
incomplete QN spectrum.

\subsection{Angles between basis functions}\label{BasAng}

In Sec.\,II we introduced covariant basis functions $\phi^m(t)$ with 
the definition $\phi^m(t)\cdot e^{i\omega_nt}=\delta_{mn}$.
If the basis functions $e^{i\omega_nt}$ were orthogonal we would
have that $\phi^n(t)$ and $e^{i\omega_nt}$ are ``parallel,''
that is,
$\phi^n(t)\propto e^{i\omega_nt}$. It is plausible
that such statements as ``this wave is dominated by oscillation at the
fundamental QN frequency'' are most meaningful if the covariant and
contravariant basis vectors are nearly ``parallel.''
To measure the extent to which 
$\phi^j(t)$ and $e^{i\omega_jt}$ fail to be proportional we can introduce
an angle $\alpha_j$ between them, defined by
\begin{equation}\label{cos-j}
  \cos(\alpha_j) = \frac{\phi^j(t) \cdot e^{i\omega_jt} }{||\phi^j(t)|| \;\; || e^{i\omega_jt} ||}  
                 = \frac{1}{\sqrt{G_{jj}G^{jj}}\ ,
}
\end{equation}
where $G^{ij}$ is the matrix inverse of $G_{ij}$.

\begin{minipage}[bth]{0.98\hsize}
\begin{minipage}[t]{0.49\hsize}
\begin{figure}[bth]
   \epsfxsize=\hsize\epsfbox{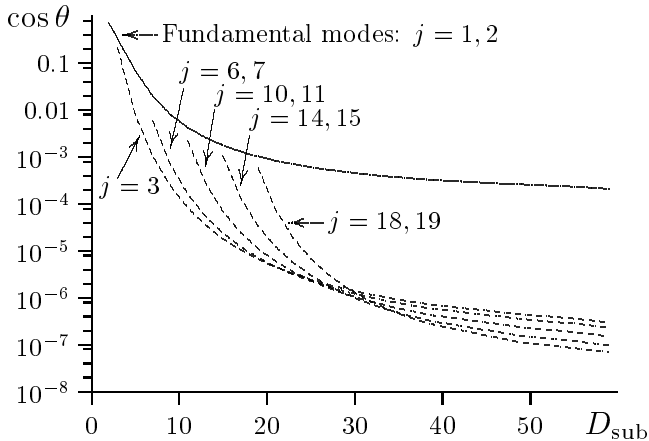}
\vspace{0.5\baselineskip}
\caption[Text to appear in the list of figures]
{Cosine of angles between several contravariant and covariant basis
vectors, as a function of the dimension of a subspace spanned by
$D_{\rm sub}$ QN modes of the \sTDP\ with $V_\delta = 10^{-6}$.}
\label{angle-tdp}
\end{figure}
\end{minipage}
\hfil
\begin{minipage}[t]{0.49\hsize}
\begin{figure}[bth]
   \epsfxsize=\hsize\epsfbox{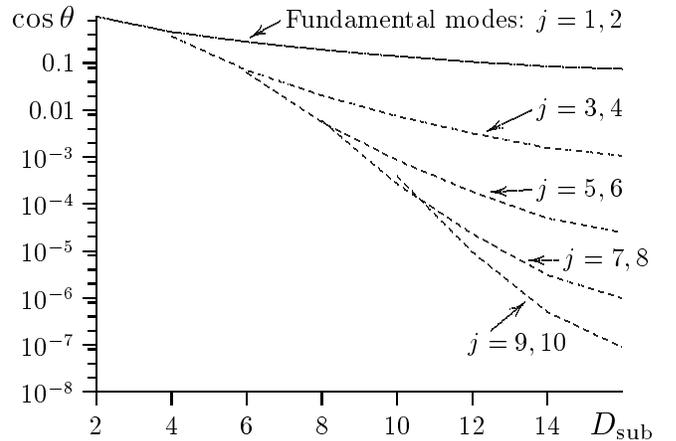}
\vspace{0.5\baselineskip}
\caption[Text to appear in the list of figures]
{Cosine of angles between several contravariant and covariant basis
vectors, as a function of the dimension of a subspace spanned by
$D_{\rm sub}$ QN modes of the Regge-Wheeler potential.} 
\label{angle-rw}
\end{figure}
\end{minipage}
\vspace{\baselineskip}
\end{minipage}
The components of the infinite matrix $G^{ij}$ cannot be computed, so
again we truncate a subspace by keeping only the first $D_{\rm sub}$
vectors, and we compute the angles in that subspace with
(\ref{cos-j}).  Results are shown in Fig.\,\ref{angle-tdp} for the
\sTDP\ with $V_\delta = 10^{-6}$. The value of $\cos{(\alpha_j)}$ is
shown for several QN modes as a function of $D_{\rm sub}$, the
dimension at which the subspace is truncated.
Figure\,\ref{angle-rw} shows
$\cos{(\alpha_j)}$ for the black hole QN spectrum. 
In Fig.~\ref{angle-tdp}, the decrease of $\cos{(\alpha_1)}$,
corresponding to the native mode, is much slower than that of the
additional modes. One might speculate that this is related to the fact
that the native mode is more characteristic of the system than the
additional ones. No such clear distinction can be seen in
Fig.~\ref{angle-rw} for the case of the Regge-Wheeler potential. 
However, the angles increase much faster for the more highly damped
modes as well, which might again indicate that the fundamental, least
damped mode is more characteristic of the system than the more
strongly damped ones.

%

\section{Acknowledgments} 
We wish to thank Ted Newman for useful discussions and numerous 
helpful suggestions on the presentations of these results.
We also thank Karel Kucha\v{r}, Wai-Mo
Suen, and John Whelan for useful discussions during the development of
this work. We thank the Deutsche Forschungsgemeinschaft for support
through SFB 382 and NATO for support through grant CRG.971092. One of
us (RHP) thanks the National Science Foundation for support under
grants PHY95-07719 and PHY97-34871.

\renewcommand{\RHd}{Quantifying QN excitations}  \markboth{\RHd}{\RHd}

\def\xd{x{\!\lower.2ex\hbox{${}_\delta$}}}

\def\psiIi{{\psi_{\rm I1}}}
\def\psiIii{{\psi_{\rm I2}}}
\def\psiIIi{{\psi_{\rm II1}}}
\def\psiIIii{{\psi_{\rm II2}}}
\def\psiIIai{{\psi_{\rm IIa1}}}
\def\psiIIaii{{\psi_{\rm IIa2}}}
\def\psiIIbi{{\psi_{\rm IIb1}}}
\def\psiIIbii{{\psi_{\rm IIb2}}}

\section*{Appendix I. finding quasinormal frequencies of the 
TDP and of the spiked TDP}

\subsection*{A. The unmodified \TDP}\label{native-qnf}

The \TDP\ is defined as
\begin{equation}
 V(x) = \cases{ 0 & $x < x_0$ \cr
                  l (l+1) / x^2 & $x \ge x_0$ \cr} ,
\end{equation}
where $l$ is an integer. 
We look only at $l = 1$, but the procedure can easily be extended for
larger values of $l$. Also, we will generally let $x_0 = 1$.

The domain of the wave equation is naturally divided into two regions:
{\labelsep=0.5cm\leftmargini=2.cm
\begin{description}
\item[I. $x < x_0$] In this region, the potential vanishes, and
therefore the wave equation has the two trivial solutions:
\begin{eqnarray}
\psiIi(x) = e^{+i\omega x} \\
\psiIii(x) = e^{-i\omega x}
\end{eqnarray}

\item[II. $x \ge x_0$] For integer values of $l$, the solutions are
given by finite sums. For $l=1$, we have
\begin{eqnarray}
\psiIIi(x) = e^{+i\omega x} (1-{1\over i\omega x}) \\
\psiIIii(x) = e^{-i\omega x} (1+{1\over i\omega x})\ .
\end{eqnarray}
\end{description}
}

Obviously, the solutions satisfying the required boundary conditions
at negative and positive infinity are
\begin{eqnarray}\label{def+-}
\psi_-(x) = \psiIi(x) & (x < x_0) \\
\psi_+(x) = \psiIIii(x) & (x \ge x_0) \ .
\end{eqnarray}

In general, of course, $\psi_-(x)$ will be a linear combination of
$\psiIIi(x)$ and $\psiIIii(x)$ for $x \ge x_0$, and $\psi_+(x)$ a combination of
$\psiIi(x)$ and $\psiIii(x)$ for $x < x_0$. A quasinormal mode is a
solution where both boundary conditions are satisfied simultaneously,
i.e. $\psi_-(x) = \psi_+(x)$. The easiest way to find out if this is
the case is to compare $\psi_-(x)$ and $\psi_+(x)$, as defined in
(\ref{def+-}), at $x = x_0$. Strictly speaking, $\psi_-(x)$ is not
defined at $x = x_0$. However, any solution of the wave equation
will have to be continuous, and have a continuous first derivative, at
$x = x_0$. It is therefore permissible to use the left limit of
$\psi_-(x)$ and of $\psi_-'(x)$ as $x \rightarrow x_{0^-}$, and
compare them with the values of $\psi_+(x)$ and $\psi_+'(x)$ at 
$x = x_0$.

The comparison is done using the Wronskian determinant of $\psi_-(x)$
and $\psi_+(x)$:

\begin{eqnarray}
W[\psi_-,\psi_+](x_0) &=& \psi_-(x_0)\psi_+'(x_0) - 
                          \psi_-'(x_0)\psi_+(x_0)   \nonumber\\
      &=& -2 i \omega -{2\over x_0} -{1 \over i \omega x_0^2}
\end{eqnarray}

Solving the equation $W[\psi_-,\psi_+](x_0) = 0$ for $\omega$ yields
the quasinormal frequencies
\begin{equation}
\omega = {\pm 1 + i \over 2 x_0}
\end{equation}
Therefore, for $l = 1$, there is only one pair of quasinormal
frequencies.

\subsection*{B. QN frequencies  of the \sTDP}\label{spiked-qnf}

We define the `spiked' potential as
\begin{equation}
\overline{V}(x) = V(x) + V_\delta  \delta (x-\xd)\ .
\end{equation}

We now have to distinguish three areas: 

\begin{description}
\item[I. $x < x_0$] This region is identical to region I for the
unmodified potential, with the same set of two solutions.

\item[IIa. $x_0 \le x \le \xd$] Again, there are two linearly
independent solutions
\begin{eqnarray}
\psiIIai(x) = e^{+i\omega x} (1-{1\over i\omega x}) \\
\psiIIaii(x) = e^{-i\omega x} (1+{1\over i\omega x})\ .
\end{eqnarray}

\item[IIb. $x > \xd$] The two linearly independent solutions are
\begin{eqnarray}
\psiIIbi(x) = e^{+i\omega x} (1-{1\over i\omega x}) \\
\psiIIbii(x) = e^{-i\omega x} (1+{1\over i\omega x})\label{psib2}\ .
\end{eqnarray}
\end{description}
In our notation the function $\psiIIbii(x)$, for example, refers to
the solution in all regions that, in region IIb, has the functional form 
shown in Eq.\ \ref{psib2}.

Due to the $\delta$ function separating regions IIa and IIb, the
solutions $\psiIIaii(x)$ and $\psiIIbii(x)$ are not the same.  Rather,
$\psiIIbii(x)$ will be a linear combination of $\psiIIai(x)$ and
$\psiIIaii(x)$ in region IIa:
\begin{equation}\label{lincomb}
\psiIIbii(x) = p_1 \psiIIai(x) + p_2 \psiIIaii(x)
\end{equation}
Once again, the solutions 
\begin{eqnarray}\label{def+-s}
\psi_-(x) = \psiIi(x) & (x < x_0) \\
\psi_+(x) = \psiIIbii(x) & (x \ge \xd)
\end{eqnarray}
satisfy the boundary conditions at negative and positive infinity.
However, in order to compare solutions at $x=x_0$, we now have to
determine the representation of $\psiIIbii(x)$ in region IIa, i.e. we
need to know the coefficients for the linear combination
(\ref{lincomb}).

These coefficients can be determined using the junction conditions for
any solution $\psi(x)$ of the wave equation across the $\delta$
function:
\begin{enumerate}
\item $\psi(x)$ must be continuous at $\xd$ 
\item  the derivative $\psi'(x)$ must have a discontinuity given by
\begin{equation}
\psi'(\xd^+) - \psi'(\xd^-) = V_\delta \psi(\xd)
\end{equation}
\end{enumerate}
The second condition is obtained by integrating the wave equation from
$x=\xd-\epsilon$ to $x=\xd+\epsilon$, letting $\epsilon \rightarrow 0$
and using the first condition. 

Using $\psiIIbii$ for $\psi'(\xd^+)$ and Eq.~(\ref{lincomb}) for
$\psi'(\xd^-)$, we can solve these conditions for $p_1$ and $p_2$. We
obtain 
\begin{eqnarray}
p_1 =     &V_\delta& 
{\left[
\psiIIaii(\xd)
\right]^2 
\over 
W_{+12}}\\
p_2 = 1 - &V_\delta& {\psiIIai(\xd)\psiIIaii(\xd) \over W_{+12}}
\end{eqnarray}

where $W_{+12} = W[\psiIIai,\psiIIaii] 
               = - 2i\omega$.

Therefore,
\begin{eqnarray}
W[\psi_-,\psi_+](x_0) &=& W[\psiIi,\psiIIbii](x_0)
=  p_1 W[\psiIi,\psiIIai](x_0) + p_2 W[\psiIi,\psiIIaii](x_0)\nonumber  \\
&=& R_1 + V_\delta \left(R_2 + R_3 e^{-2i\omega L} \right)\label{sTDPWr}  ,
\end{eqnarray}
where $L = \xd - x_0$, and 
$W[\psiIi,\psiIIai](x_0) = e^{2i\omega x_0} /(i \omega x_0^2)$, 
and thus 
\begin{eqnarray}
R_1 &=&   W[\psiIi,\psiIIaii](x_0) 
     = - 2 i \omega - {2\over x_0} - {1 \over i \omega x_0^2}\label{R1}  \\
R_2 &=& - {\psiIIai(\xd)\psiIIaii(\xd) \over W_{+12} } W[\psiIi,\psiIIaii](x_0)
    = -\left(1 - {1\over (i\omega \xd)^2}\right)\left(1+\frac{1}{i \omega x_0}
+\frac{1}{2(i \omega x_0)^2}
\right) \label{R2}\\
R_3 &=& e^{2i\omega L} {\psiIIaii(\xd)^2 \over W_{+12}} W[\psiIi,\psiIIai](x_0)
    = - {1\over 2 (i\omega x_0)^2} \left(1 + {1\over i\omega \xd}\right)^2 \label{R3}
\end{eqnarray}

The quasinormal frequencies of the \sTDP can now be computed
numerically by searching for roots of the equation
\begin{equation}\label{Eqqnf}
W[\psi_-,\psi_+](x_0)=0\ .
\end{equation}

\subsection*{C. Asymptotic approximation for the QN frequencies of the
\sTDP}

It is possible to find an asymptotic formula for the
QN frequencies under the assumption that the absolute value of the frequency
becomes large. We start by assuming that in (\ref{Eqqnf}), we have $|\omega x_0|\gg1$
and $|\omega x_\delta|\gg1$. The condition for QN frequencies can then be written
\begin{equation}
  \label{order}
  2i\omega x_0 +{\cal O}([\omega x)]^0)
+V_\delta\left[1+{\cal O}([\omega x)]^{-1})
+\frac{e^{-2i\omega L}}{2(i\omega x_0)^2}
\left(
1+{\cal O}([\omega x)]^{-1})
\right)
\right]=0\ ,
\end{equation}
where $x$ is the minimum of $x_0$ and $x_\delta$.
It is clear that for $|\omega x|\gg 1$ there can be solutions only
if 
\begin{equation}
2 i \omega + V_\delta {1 \over 2 (i\omega x_0)^2} 
                 e^{-2i\omega L}   \approx0\ ,
\end{equation}
and we use this approximation to find an iterative solution for the
QN frequencies. We start by taking the cube root of
\begin{equation}
 e^{-2i\omega L} = - {4 (i\omega)^3 x_0^2 \over V_\delta}
\end{equation}
to write
\begin{equation}\label{EqAs}
 e^{-{2\over 3} i \omega_R L} \,
   e^{ {2\over 3} \omega_I L}   \, e^{i\Delta} = 
   - \left({4 x_0^2 \over V_\delta}\right)^{1\over 3} \; i \omega
   \equiv -A \, i \omega  ,
\end{equation}
where $\Delta = j {2\over 3} \pi, \; j = 0,1,2$.

Taking the absolute values on both sides gives
\begin{equation}\label{EqAbs}
    e^{ {2\over 3} \omega_I L} = A |\omega|
\end{equation}
With $\omega\equiv \omega_R+i\omega_I$ this last relation already
shows that $\omega_I \ll |\omega| \approx \omega_R$ is required for a QN
frequency.

Using (\ref{EqAbs}) to rewrite (\ref{EqAs}) we find
\def\gam{{2\over 3}L}
\begin{equation}
\cos(\gam\omega_R + \Delta) - i \sin(\gam\omega_R + \Delta) = 
  -i{\omega \over |\omega|}
\end{equation}
We now make the approximation $\omega_I \approx 0$, i.e. 
$\omega \approx \omega_R \approx |\omega|$, leading to 
\begin{equation}
\cos(\gam\omega_R + \Delta) = 0, \quad
\sin(\gam\omega_R + \Delta) = 1 
\end{equation}
\begin{equation}
\gam\omega_R = {\pi\over 2} + 2 N \pi - {2 \over 3} j \pi, \quad
   N = 0,1,... ;  \; j = 0,1,2
\end{equation}
or equivalently
\begin{equation}\label{asym_R}
(\omega_R)_n = {1\over L} ({3 \pi\over 4} + n \pi), \quad  n = 0,1,...
\end{equation}
An approximation for $\omega_I$ is then obtained by using (\ref{EqAbs}):
\begin{equation}\label{asym_I}
(\omega_I)_n = {3 \over 2} {1 \over L} (\ln A + \ln \omega_R)
\end{equation}

The approximate solutions of (\ref{asym_R}) and (\ref{asym_I}) 
can now be iteratively improved. We rewrite (\ref{Eqqnf}) as
\begin{equation}\label{Eqrqnf}
e^{-2i\omega L} = - {R_1 + V_\delta R_2 \over V_\delta R_3} 
   =: R(\omega) = R(\omega_R,\omega_I)
\end{equation}
and take absolute values to get
\begin{equation}
  e^{2\omega_I L} = | R | 
\end{equation}
\begin{equation}
\omega_I = {\ln|R| \over 2 L}  
\end{equation}
With these inserted in (\ref{Eqrqnf}) we arrive at 
\begin{equation}
  \cos(2\omega_R L) = - {\Re(R) \over |R|}, \quad
   \sin(2\omega_R L) =   {\Im(R) \over |R|}\ .
\end{equation}
This can be used as  an iterative method to find the $p^{\rm th}$ iteration from the 
$p-1^{\rm th}$ approximation, as follows:
\begin{equation}
  \cos(2(\omega_R)^p L) = - {\Re(R((\omega_R)^{p-1},(\omega_I)^{p-1})) 
                              \over |R((\omega_R)^{p-1},(\omega_I)^{p-1})|}, 
   \quad
   \sin(2(\omega_R)^p L) =   {\Im(R(...)) \over |R(...)|}
\end{equation}
and
\begin{equation}
(\omega_I)^p = {\ln|R((\omega_R)^p,(\omega_I)^{p-1})| \over 2 L}  
\end{equation}
This iterative solution can be started with {\em any} value of $n$ in
the zeroth approximation of (\ref{asym_R}) and (\ref{asym_I}), and the
iteration converges to the exact solution. The iteration cannot be used to
find the native QN frequency of the smooth TDP since $V_\delta=0$ in that case.

%
\renewcommand{\RHd}{Quantifying QN excitations}  \markboth{\RHd}{\RHd}

\section*{Appendix II. Proof of convergence}

A proof is given here of the convergence of the sum of quasinormal excitations
for the spiked TDP under approprite restrictions on the Cauchy data.
To do this we start by defining the integral
\begin{equation}\label{defInt}
{\cal I}(d_1)\equiv\int F(s)\,ds\equiv\frac{1}{2\pi i}
\int \frac{s^2\phi(s)e^{s(d_1+d_2)}}{\Delta(s)}\,ds\ .
\end{equation}
Here $\Delta(s)$ is defined to be
\begin{equation}
  \label{defDelta}
\Delta(s)\equiv P_1(s)+e^{2sL}s^3P_2(s)\ ,  
\end{equation}
in which $P_1$ and $P_2$ are polynomials of finite order in $1/s$:
\begin{equation}
  \label{defPs}
P_1(s)\equiv A_1+B_1/s+\cdots \quad\quad   
P_2(s)\equiv A_2+B_2/s+\cdots \quad\quad   \ .
\end{equation}
The path of integration in the complex $s$ plane is along the
vertical axis, from $-i\infty$ to $+i\infty$.  
For $\Re(s)\le0$ the function $\phi(s)$ is required to have
the property that for some real constants $K_\phi$ and $p$
\begin{equation}\label{condition1}
\left|\phi(s)\right|<\frac{K_\phi}{|s|^p}\ .
\end{equation}
(This condition will be related below to restrictions on acceptable
Cauchy data.)  The constants appearing in (\ref{defInt}) are taken to
satisfy the following conditions: (i) The ratio $A_1/A_2$ is real and
positive. (ii) $L, d_1$ and $d_2$ are real and nonnegative, and 
$2L>d_2$. (iii) The
constant $p$ appearing in (\ref{condition1}) must be large enough that
\begin{equation}\label{condition2}
p+\frac{3d_2}{2L}-2>0\ .
\end{equation}

The roots of $\Delta$ are denoted $s_k$. (They represent, of course,
the QN frequencies according to the usual correspondence
$s\leftrightarrow i\omega$.) . Since they must occur in complex
conjugate pairs it is convenient here to use the notation
$s_{1},s_{-1},s_{2},s_{-2},\cdots$ with $s_{\pm k}$ indicating the
corresponding roots with positive and negative imaginary parts.
\begin{figure}[bth]
   \hspace*{2in}\epsfxsize=.25\hsize\epsfbox{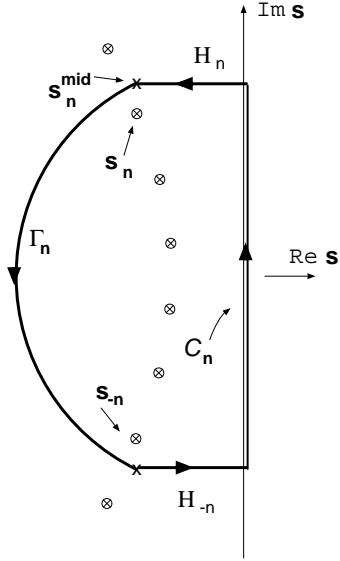}
\caption[Text to appear in the list of figures]
{Contour for proving convergence for the \sTDP\ oscillations.} 
\label{fig:proof}
\end{figure}

What we will prove is that under the conditions stated above
${\cal I}(d_1)$ can be written as
\begin{equation}\label{2bproven}
{\cal I}(d_1)=\sum_{k=-\infty}^{k=+\infty}a^{(k)}e^{s_kd_1}
\end{equation}
where 
\begin{equation}\label{coefvals}
  a^{(k)}\equiv
  s_k^2\phi(s_k)e^{s_kd_2}/\left(d\Delta/ds\right)|_{s=s_k}\ .
\end{equation}
The convergence of the sum in (\ref{2bproven}) is not uniform at
$d_1\rightarrow0$, but is uniform for any interval of $d_1$ bounded
away from 0.  We give the details of the relationship to the physical
problem of Sec.~III after we prove the main result above.  The proof
is organized with a set of lemmas.


{\em Lemma 1:} Let $s^{mid}_n=s_n+i\pi/2L$ and let $H_n$ be the
horizontal path (as shown in Fig.\,\ref{fig:proof}) in the $s$ plane
from the imaginary $s$ axis to the point $s^{mid}_n$. Let
\begin{equation}
{\cal I}_{H_n}\equiv\int_{H_n}F(s)\,ds
\end{equation}
be the integral of $F(s)$ on this path, then $N$ can be chosen so that
for $n>N$
\begin{equation}
\left|{\cal I}_{H_n}\right|\le{\rm const}\times n^{2-p-(3d_2/2L)}\ .
\end{equation}

{\em Proof:} The discussion of the roots of the \sTDP\ Wronskian
in Appendix I can be applied to the roots of $\Delta$. For large
$n$, the $n$ dependence of the roots takes the form
\begin{eqnarray}\label{imsres}
\Im(s_n)&=&n\pi/L+{\rm const.}+\cdots\\
\Re(s_n)&=&-(3/2L)\ln(n)+{\rm const.}+\cdots\,.
\end{eqnarray}
It follows that a constant can be chosen so that
$|s_n|>{\rm const}\times n$, and hence $|\phi(s)|<2\pi K/n^p$, for some
constant $K$ independent of $n$.  Since $\Re(s)\le0$ on $H_n$ and
$d_1$ is nonnegative, we have that $|e^{sd_1}|\le1$.  With
$s=s_n+i\pi/2L+\sigma$ and with $\sigma$ running from $-\Re(s_n)$ to 0 on
the path $H_n$, the integral must satisfy
\begin{equation}
|{\cal I}_{H_n}|\le\frac{K}{n^{p-2}}\left| \int_0^{-\Re(s_n)}\frac{d\sigma}{D(s)}\right|\ ,
\end{equation}
where $D(s)\equiv e^{-sd_2}\Delta(s)$.
On $H_n$ we have that
\begin{equation}
e^{2Ls}=-e^{2Ls_n}e^{2L\sigma}
\end{equation}
and $D(s)$ can be written as
\begin{equation}\label{defD}
D(s)=P_1(s_n)e^{-i\Im(s)d_2}\left[e^{-\Re(s)d_2}{\cal R}_1
+e^{2L\sigma}e^{-\Re(s)d_2}{\cal R}_2
\right]
\end{equation}
where 
\begin{equation}
{\cal R}_1=\frac{P_1(s)}{P_1(s_n)}\quad\quad
{\cal R}_2=\frac{s^3P_2(s)}{s_n^3P_2(s_n)}    \ .
\end{equation}
For $n$ larger than some $N$ we can make both ${\cal R}_1$ and ${\cal
  R}_2$ arbitrarily close to unity. For $n>N$ it follows that the magnitude
of the sum in square brackets in (\ref{defD}) must be larger than the
second term, and hence
\begin{equation}
|D(s)|>|P_1(s_n)|
e^{2L\sigma}e^{-\Re(s)d_2}|{\cal R}_2|
\end{equation}
By choosing $N$ large enough we can make 
$|P_1(s_n)|$ 
and $|{\cal R}_2|$ larger than $n$-independent constants,
so that 
\begin{equation}
  |D(s)|>{\rm const}\times e^{2L\sigma}e^{-\Re(s)d_2}
={\rm const}\times e^{2L\sigma}e^{-\Re(s_n)d_2}e^{-\sigma d_2}\ .
\end{equation}
We have then that 
\begin{eqnarray}
|{\cal I}_{H_n}|&\le&\frac{K'}{n^{p-2}}e^{\Re(s_n)d_2}\left| \int_0^{-\Re(s_n)}
e^{\sigma(d_2-2L)}
d\sigma
\right|\nonumber\\
&<&\frac{K'}{n^{p-2}}e^{\Re(s_n)d_2}\left|\frac{1}{2L-d_2}
\right|\ .
\end{eqnarray}
It follows from (\ref{imsres}) that we can choose a constant so that
$|e^{\Re(s_n)d_2}|<{\rm const.}/n^{(3d_2/2L)}$, 
and therefore that
\begin{equation}
|{\cal I}_{H_n}|\le\frac{{\rm const}}{n^{p+(3d_2/2L)-2}}\ ,
\end{equation}
which was to be proven. A similar proof shows that the same result
applies to $|{\cal I}_{H_{-n}}|$.


{\em Lemma 2:} On the   arc $\Gamma_n$     from 
$s^{mid}_n$ to $s^{mid}_{-n}$ 
with $|s|=|s^{mid}_n|$, 
the magnitude of $\Delta(s)$
is larger than some constant that is independent of $n$.

{\em Proof:}
We write 
\begin{equation}
\Delta(s)=P_1(s)\left[1+\Phi(s)\right]
\end{equation}
where
\begin{equation}
\Phi(s)\equiv e^{2sL}s^3P_2(s)/P_1(s)\ .
\end{equation}
We must show that $|1+\Phi|$ is bounded from below by an $n-$independent
constant.
We recall that $s^{mid}_n=s_n+i\pi/2L$ and that $\Phi(s_n)=-1$, so
that
\begin{equation}
\Phi(s^{mid}_n)\equiv 
\left(\frac{s^{mid}_n}{s_n}\right)^3
\frac{P_2(s^{mid}_n)}{P_2(s_n)}\,
\frac{P_1(s_n)}{P_1(s^{mid}_n)}\ .
\end{equation}
For $n$ larger than some $N_{\rm min}$ we have that 
$\Phi(s^{mid}_n)$ is arbitrarily close to unity, so that
\begin{equation}
\Phi(s^{mid}_n)=1+\lambda,
\end{equation}
and $N_{\rm min}$ can be chosen large enough 
to make $|\lambda|$ arbitrarily small.

On the arc $\Gamma_n$ we write $\Phi(s)$ as 
\begin{equation}
  \Phi(s)=e^{2sL}s^3\left(A_2/A_1\right)\left[
1+\rho(s)
\right]\ .
\end{equation}
By choosing $N_{\rm min}$ sufficiently large, we can bound the
magnitude of $\rho$ to be less than an arbitrarily small
$n$-independent constant.  For $s$ on the arc $\Gamma_n$, we now write
$s$ as
\begin{equation}
  \label{deftheta}
  s=R_nie^{i\theta}\ ,
\end{equation}
where $R_n\equiv|s^{mid}_n|$ and $\theta$ is the counterclockwise
angle from the positive imaginary $s$ axis to the point $s$ on
$\Gamma_n$. We denote by $\theta_n$ the angle to $s^{mid}_n$, so that
$s^{mid}_n\equiv R_nie^{i\theta_n}$ and we write
$\theta\equiv\theta_n+\delta\theta$. In terms of this notation we have
\begin{equation}
  \Phi(s)=F_0F_1(\delta\theta)F_2(\delta\theta)\ ,
\end{equation}
where 
\begin{eqnarray}
F_0&\equiv&(1+\lambda)(1+\rho(s))/
(1+\rho(s^{mid}_n))\label{defF0}\\
   F_1&\equiv& e^{2iLR_n\left[
\cos(\theta_n+\delta\theta)-\cos(\theta_n)
\right]}\ e^{3i\delta\theta}\label{defF1}\\
   F_2&\equiv& e^{2LR_n\left[
\sin(\theta_n)-\sin(\theta_n+\delta\theta)
\right]}\label{defF2}\ .
\end{eqnarray}
The complex phase of $\Phi(s)$ is near zero at $s=s^{mid}_n$, and decreases
as $s$ moves counterclockwise along the arc $\Gamma_n$. 
We use the expressions above to find at what
value $\delta\theta^*$
of $\delta\theta$ the phase of $\Phi$ first becomes $-\pi/2$. We note
that  $|F_1|=1$ and on the top half of the arc $F_2\leq1$.  The 
value of $\delta\theta^*$ is given by
\begin{equation}
  \label{delphieq}
  {2LR_n\left[
\cos(\theta_n+\delta\theta^*)-\cos(\theta_n)
\right]}+3\delta\theta^*
+
\zeta(\theta_n+\delta\theta^*)=-\pi/2\ ,
\end{equation}
where $\zeta(\theta)$ is the phase of $F_0$. 
We note that 
\begin{displaymath}
|\cos(\theta_n+\delta\theta^*)-\cos(\theta_n)|>
\sin(\theta_n)\,\delta\theta^*\ .
\end{displaymath}
From (\ref{imsres}) we know that $\sin\theta_n$ decreases with $n$ as
$\ln(n)/n$, and $R_n$ increases as $n$. Thus the first term
in (\ref{delphieq}) 
is larger than the second by a factor that increases
as $\ln(n)$. The third term, the $\zeta$ term, decreases with
increasing $n$, and we can bound it to be smaller than an arbitrarily
small constant by choosing $N_{\rm min}$ sufficiently large. From
these considerations it follows that we can choose $N_{\rm min}$ large
enough that the magnitude of the first term in
(\ref{delphieq}) is larger than, say, 2/3 of the magnitude of the left 
hand side, and hence $\delta\theta^*$ satisfies
\begin{equation}
  {2LR_n\left[\cos(\theta_n)-
\cos(\theta_n+\delta\theta^*)
\right]}>\pi/3\ .
\end{equation}
Let us also take $N_{\rm min}$ large enough so that
$\theta_n+\delta\theta^*<\pi/4$. In that case we have
\begin{equation}
  \label{trigineq}
  \frac{
\sin(\theta_n+\delta\theta^*)-\sin(\theta_n)
}
{
\cos(\theta_n)-\cos(\theta_n+\delta\theta^*)
}
>1\ .
\end{equation}
From this it follows that
$2LR_n[\sin(\theta_n+\delta\theta^*)-\sin(\theta_n)]>\pi/3$, and
  hence $F_2<e^{-\pi/3}$. Since the deviation of $|F_0|$ from unity is
  arbitrarily small, let us use $|F_0|<e^{\pi/12}$ and conclude that
  $|\Phi|<e^{-\pi/4}$ at the point along $\Gamma_n$ at which $\Phi$
  first becomes purely imaginary. As $\delta\theta$ increases, the
  magnitude of $\Phi$ continues to decrease. It follows that for every
  point along the top of the arc
\begin{equation}
  \label{bound}
  |1+\Phi|>1-e^{-\pi/4}\ .
\end{equation}
A similar analysis starting at $s_{-n}$ shows that (\ref{bound})
holds also for the bottom half of the arc.


{\em Lemma 3:} On the   arc $\Gamma_n$     from 
$s^{mid}_n$ to $s^{mid}_{-n}$ 
with $|s|=|s^{mid}_n|$, the integral
\begin{equation}
{\cal I}_{\Gamma_n}(d_1)
\equiv\frac{1}{2\pi i}
\int \frac{s^2\phi(s)e^{s(d_1+d_2)}}{\Delta(s)}\,ds
\end{equation}
satisfies, 
\begin{equation}
|{\cal I}_{\Gamma_n}(d_1)|<\frac{\rm const}{n^{p+(3d_2/L)-2}}\ ,
\end{equation}
where the constant is independent of $n$.

{\em Proof:}
On $\Gamma_n$ we have that 
$\Re(s)\le\Re(s_n)$, so that 
\begin{equation}\label{esd2}
|e^{sd_2}|\le|e^{\Re(s_n)d_2}| 
\end{equation}
and, for some constant, the right hand side of
(\ref{esd2})
is less than ${\rm
  const}/n^{(3d_2/2L)}$.  We have seen that $|\Delta|$ is bounded from
below on $\Gamma_n$. With the bound on $|\phi(s)|$ from
(\ref{condition1}), we have then that
\begin{equation}
|{\cal I}_{\Gamma_n}(d_1)|< \frac{\rm const}{n^{p+(3d_2/2L)-2}}\int_{\Gamma_n}
e^{\Re(s)d_1}|ds|\ .
\end{equation}
Since the integrand is everywhere nonnegative, we have that
\begin{equation}
\int_{\Gamma_n}
e^{\Re(s)d_1}|ds|<\int_{\rm arc}
e^{\Re(s)d_1}|ds|\ ,
\end{equation}
where the arc is the half circle with $|s|=|s^{mid}_n|=R_n$ in the left
half plane.  But, the integral along the half circle is 
\begin{eqnarray}
  \int_{\rm arc}
e^{\Re(s)d_1}|ds| &=&R_n\int_{0}^{\pi}e^{-d_1R_n\sin{\theta}}d\theta
=2R_n\int_{0}^{\pi/2}e^{-d_1R_n\sin{\theta}}d\theta\nonumber\\
&<&R_n\int_{0}^{\pi/2}e^{-2d_1R_n{\theta}/\pi}d\theta=\frac{\pi}{d_1}\left(
1-e^{-d_1R_n}
\right)\ .
\end{eqnarray}
This completes the proof of
the lemma.


{\em Proof of main result:}
We define $C_n$ as the integration path on the imaginary $s$ axis
from $i\Im(s_{-n})-i\pi/2L$ to $i\Im(s_{n})+i\pi/2L$, and we define
\begin{equation}
{\cal I}_n(d_1)\equiv\int_{C_n}F(s)\,ds\ .
\end{equation}
We let ${\cal I}_{\rm n,closed}(d_1)$  be the integral on the closed path
consisting of 
$C_n$, of $\Gamma_n$, of $H_n$ and of $H_{-n}$ traced backwards. 
From the lemmas above we have 
\begin{eqnarray}
  |{\cal I}_n(d_1)-{\cal I}_{\rm n,closed}(d_1)|&<|{\cal
    I}_{\Gamma_n}(d_1)|+|{\cal I}_{H_n}(d_1)|+|{\cal I}_{H_{-n}}(d_1)|
&<\frac{{\rm const}}{n^{p+(3d_2/2L)-2}}\ .
\end{eqnarray}
The integral on the closed path is $2i\pi$ times the sum of the residues inside
the path,
\begin{equation}
{\cal I}_{\rm n,closed}(d_1)=\sum_{-n}^{n}a^{(k)}e^{s_kd_1}\ ,
\end{equation}
where $a^{(k)}$ is the residue of $s^2\phi(s)e^{sd_2}/\Delta$ at
$s=s_k$.  Since the only singularities of the integrand in the finite
$s$ plane are simple poles at the roots of $\Delta(s)$, these
$a^{(k)}$ coefficients are those defined in (\ref{coefvals}).
We have then 
\begin{equation}
  |{\cal I}_n(d_1)-\sum_{-n}^{n}a^{(k)}e^{s_kd_1}|<
\frac{{\rm const}}{n^{p+(3d_2/2L)-2}}
\end{equation}
and  our  main result follows from the fact that 
${\cal I}(d_1)$ is the limit of 
${\cal I}_n(d_1)$
as $n\rightarrow\infty$.

We now turn to the role played by the Cauchy data. In the Green
function solution for the waveform [see the discussion following
(\ref{Laplwave})], a function of $s$ appears representing the integral
of the product of $f_{-}(s,x)$ and the combination
$J(x,s)\equiv-\dot{\psi}_0(x)-s\psi_0(x)$. In the case of the TDP or
spiked TDP, $f_{-}(s,x)=e^{sx}$.  Let us suppose that the support of
the Cauchy data is confined to the region $x_2< x< x_1$ The Cauchy
data then enters the $s$ integral through the function
\begin{equation}
{\cal J}(s)\equiv \int_{x_2}^{x_1} J(x,s)e^{sx}\,dx\ .
\end{equation}
If the initial waveform
$\psi_0(x)$
satisfies
\begin{equation}\label{psismooth}
\left|\frac{d^{p+1}}{dx^{p+1}}\psi_0(x)\right|<b_0\ ,
\end{equation}
then from integration by parts, we have
\begin{eqnarray}
\int_{x_2}^{x_1}e^{sx}\psi_0(x)\,dx&=-\frac{1}{s}\int_{x_2}^{x_1}e^{sx}\frac{d}{dx}[\psi_0(x)]\,dx\\
&=\cdots\pm\frac{1}{s^{p+1}}\int_{x_2}^{x_1}e^{sx}\frac{d^{p+1}}{dx^{p+1}}[\psi_0(x)]\,dx\\
&=\cdots\pm\frac{1}{s^{p+1}}e^{sx_2}\int_{x_2}^{x_1}e^{s(x-x_2)}\frac{d^{p+1}}{dx^{p+1}}[\psi_0(x)]\,dx\ .
\end{eqnarray}
For $\Re(s)\le0$ the factor $e^{s(x-x_2)}$ in the last integral is $\le1$, so that
\begin{equation}
\left|e^{-sx_2}\int_{x_2}^{x_1}e^{sx}\psi_0(x)\,dx\right|<\frac{|b_0(x_1-x_2)|}{|s|^{p+1}}\ .
\end{equation}
If in addition to the constraint in (\ref{psismooth}) we have that
the $p^{\rm th}$ derivative of $\dot{\psi}_0(x)$ is bounded,
then by a very similar argument we can show that
$\phi(s)$, defined as $e^{-sx_2}{\cal J}(s)$, 
satisfies
\begin{equation}
|\phi(s)|<\frac{\rm const}{|s|^p}\ .
\end{equation}

We can now apply the above mathematical results to the Green function integral from Sec.~III.
The waveform is given by the following integral along the imaginary $s$ axis:
\begin{equation}
\psi(t,x)=\frac{1}{2i\pi}\int \frac{e^{s(t-x)}{\cal J}(s)}{W(s)}\,ds
\end{equation}
where $W(s)$ is given in 
(\ref{sTDPWr}) -- (\ref{R3}) and has
has the form
$W(s)=s^{-2}e^{-2sL}\Delta(s)$ in which $\Delta(s)$ is a special 
case of (\ref{defDelta}) and
(\ref{defPs}). We can therefore rewrite the solution as
\begin{equation}\label{proofpsi}
  \psi(t,x)=\frac{1}{2i\pi}\int
  \frac{s^2e^{s(t-x+x_2+2L)}\phi(s)}{\Delta(s)}\,ds\ .
\end{equation}


The above proof requires that $p,d_1,d_2$ and $L$ satisfy the
inequalities that follow (\ref{condition1}). The details of the proof
show that the rate of convergence depends on these parameters. In
particular, on the value of $\gamma\equiv p+(3d_2/2L)-2$. A small
value of this parameter means slow convergence. We can now relate the
details of the proof to the examples presented in Sec.\,III, and
examine the interesting nature of the convergence of the series given
there.  We start by noting that a straightforward computation of
$\phi(s)\equiv e^{-sx_2}{\cal J}(s)$, for the Cauchy data of
(\ref{inidat}), shows that $p=1$. Comparing (\ref{defInt}) and
(\ref{proofpsi}) we see that
\begin{eqnarray}
  \label{d1d2}
  d_1+d_2&=&t-x+x_2+2L\nonumber\\
&=&t-x+x_1-x_1+x_2+2L\ .
\end{eqnarray}
An ``obvious'' choice is to take $d_1=t-x+x_1$, the retarded time from
the start of the reception of signals from the Cauchy data. At any $x$
this is the equivalent of $t-t_{\rm min}$. With this choice we are
left with $d_2=x_2-x_1+2L$. (Note that $2L-d_2=x_1-x_2$ is
positive, as required in the proof.) The value of $\gamma$ for this
choice is given by 
\begin{equation}
  \label{gammaeq}
\gamma\equiv p+\frac{3d_2}{2L}-2
=2-3\frac{x_1-x_2}{2L}\ .
  \end{equation}
  In our examples, we choose $x_1-x_2$, the range of support of the
  Cauchy data, to be 6 and we have $L=9$, so $\gamma$ is unity.
  Suppose, though, that we had chosen $x_1=1$ (as in our standard sine
  wave data of Sec.\,III) but had decreased the value of $x_2$ below
  our standard choice -5. The details of the proof show that convergence
  would require more terms for a given level of accuracy and that
  the series would fail to converge for $x_2\le11$. 
    This limit can be extended if we use initial data with one or more
    continuous derivatives, i.e. if $p > 1$. 
  This rather
  unusual feature was, in fact, exactly what was observed in numerical
  experiments.

  We point out next that Lemma 3 shows that convergence is not uniform
  in $d_1$.  As $d_1$ gets smaller, more terms in the series are
  needed. With our choice of $d_1$ to be $t-t_{\rm min}$, this implies
  that the convergence of our QN series in (\ref{2bproven}) is not
  uniform as $t-t_{\rm min}\rightarrow0$, contradicting our claims of uniform
  convergence made following (\ref{compdef}).  But note that we can
  choose $d_1\equiv t-t_{\rm min}+1$, so that convergence {\em is}
  uniform for $t>t_{\rm min}$. In this case we have
  $\gamma=2-3(x_1-x_2+1)/2L$. For both our sine wave Cauchy data this
  $\gamma$ has a numerical value of $5/6$, and the series is
  convergent. It is clear that that there is an interaction between
  the allowed range of the Cauchy data, and the range of $t$ for which
  the QN series converges. By moving the left edge of the support of
  the Cauchy data to the left by some amount $\delta$, we increase by
  $3\delta/(2L)$ the value of $t$ at which the series first converges.
It should also be noted that 
the dependence on $d_1$ explains why the QN series converge more quickly 
at early times than at late, a feature evident in
Fig.\,\ref{ms_E6}.
\renewcommand{\RHd}{Quantifying QN excitations}


\end{document}